\documentclass[preprint,12pt]{elsarticle}

\usepackage{amssymb}
\usepackage{amsmath,graphicx,psfrag,epsf,xcolor,url}
\usepackage{enumerate}
\usepackage{natbib}
\usepackage{subcaption}
\usepackage{algorithm}
\usepackage{algpseudocode}
\newtheorem{definition}{Definition}

\newtheorem{theorem}{Theorem}

\newtheorem{corollary}{Corollary}
\allowdisplaybreaks
\numberwithin{equation}{section}
\numberwithin{figure}{section}
\numberwithin{table}{section}
\numberwithin{theorem}{section}
\numberwithin{lemma}{section}
\numberwithin{proposition}{section}
\numberwithin{corollary}{section}
\journal{Physica A}

\begin{document}

\begin{frontmatter}



\title{Measuring Interlayer Dependence of Large Degrees in Multilayer Inhomogeneous Random Graphs} %% Article title

\author[label1]{Zhuoye Han}
\affiliation[label1]{organization={School of Mathematical Sciences, Fudan University},
	%%addressline={},
	city={Shanghai},
	postcode={200433},
	%%state={},
	country={China}}
\author[label2,label3]{Tiandong Wang}
\affiliation[label2]{organization={Shanghai Center for Mathematical Sciences, Fudan University},
	%%addressline={},
	city={Shanghai},
	postcode={200433},
	%%state={},
	country={China}}         
\affiliation[label3]{organization={Shanghai Academy of Artificial Intelligence for Science},
	%%addressline={},
	city={Shanghai},
	postcode={200003},
	%%state={},
	country={China}}   


%%\author{} %% Author name

%%\affiliation{organization={},%Department and Organization
	%%           city={},

%% Abstract
\begin{abstract}
Accurately capturing interlayer dependence is essential for understanding the structure of complex multilayer networks. We propose an upper tail dependence estimator specifically designed for multilayer networks, leveraging multilayer inhomogeneous random graphs and multivariate regular variation to model extremal dependence. We establish the consistency of the estimator and demonstrate its practical effectiveness through real-data analysis of Reddit. Our findings reveal how financial market dynamics influence user interactions in the BitcoinMarkets subreddit and how seasonal trends shape engagement in sports-related subreddits. This work provides a rigorous and practical tool for quantifying extremal dependence across network layers, offering valuable insights into risk propagation and interaction patterns in multilayer systems.
\end{abstract}

\begin{graphicalabstract}
	\includegraphics[width=\textwidth]{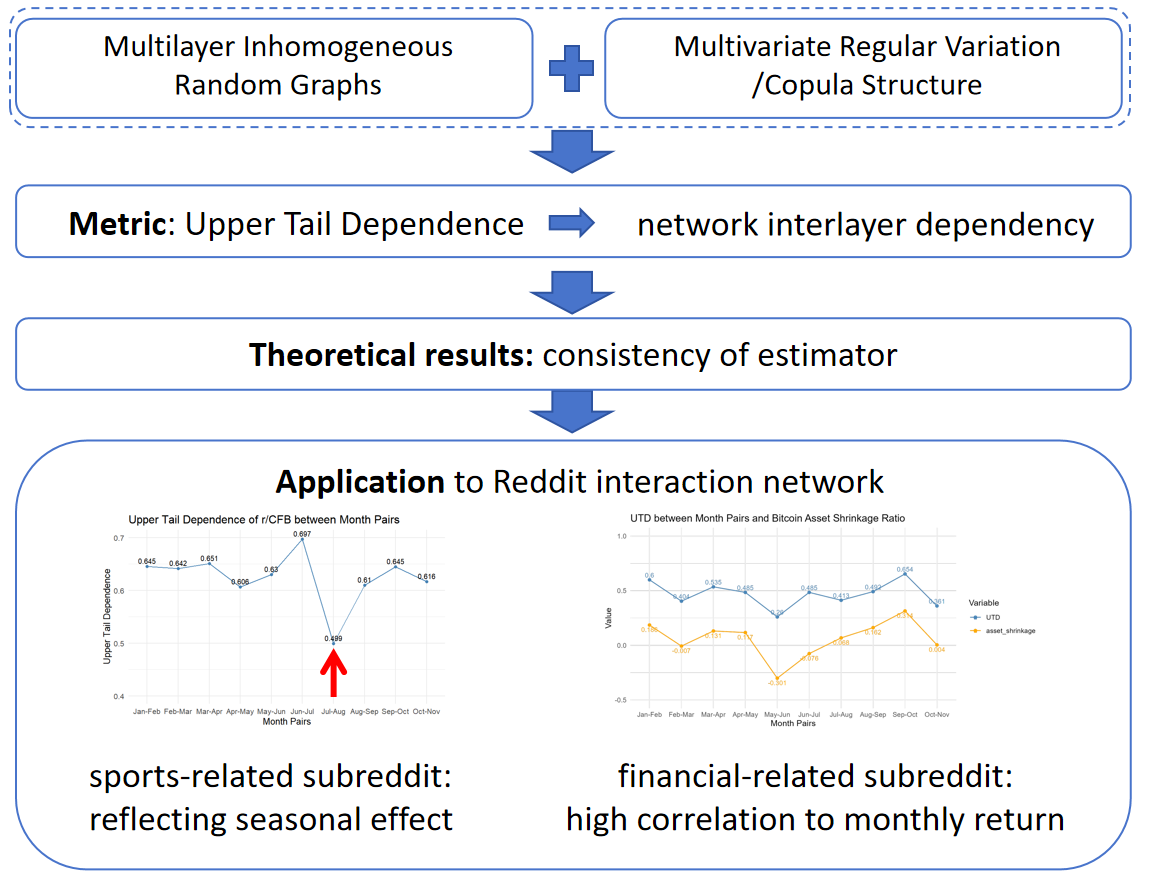}
\end{graphicalabstract}

\begin{highlights}
\item Introduce the upper tail dependence (UTD) metric based on multilayer inhomogeneous random graphs.
\item Build the estimator and prove its consistency under mild conditions.
\item Validate estimators via simulations with Gumbel copula and multivariate regular variation structure.
\item Apply UTD to Reddit interaction network: uncover market-behavior links in BitcoinMarkets subreddit and seasonal impacts on sports-related subreddits.
\end{highlights}

%% Keywords

\end{frontmatter}


\section{Introduction}
\label{sec:intro}

Understanding interdependence in multilayer networks is crucial for analyzing complex systems across domains such as social networks, finance, and biology. These networks capture diverse interactions, where disruptions in one layer can cascade through others, leading to systemic failures. For example, transport disruptions can spread across modes, and financial correlations can propagate risks. Accurately characterizing these dependencies is essential for risk assessment, decision-making, and policy development.

Traditional dependence measures, such as Pearson's and Kendall's correlation coefficients, have been widely used to quantify relationships in single-layer networks \citep{xie2022systemic}, but do not capture complex dependencies between layers in multilayer networks. To address this, advanced methods have been explored: \cite{si2019brain} used mutual information to analyze changes in brain connectivity related to Alzheimer's, \cite{danziger2019dynamic} modeled node features to define cross-layer dependencies, \cite{min2014network} examined interlayer interactions on multiplex hypergraphs, and \cite{tewarie2021interlayer} proposed a multilayer network approach to reconstruct interlayer connectivity in neurophysiological networks. While these studies provide insights into multilayer dependencies, they do not specifically address multivariate extreme events, which are crucial for understanding risk propagation and system stability.

To fill this gap, multivariate regular variation (MRV) and copula functions offer powerful tools for modeling extremal dependence. MRV has been widely applied in finance and risk management \citep{cai2011estimation,das2018risk} to analyze extreme risk propagation, while copula functions, particularly the Gumbel copula, provide a flexible approach to capture the upper tail dependence. For example, \citep{lee2013data} used copulas for bivariate drought analysis, \citep{das2023systemic} applied them to study risk contagion in financial networks, and \cite{joe2011tail} combined copulas with MRV to examine tail dependence structures in non-network settings. Despite their success in other domains, these methods remain underutilized in multilayer network analysis.

To bridge this gap, we propose an upper tail dependence (UTD) estimator based on multilayer inhomogeneous random graphs (MIRG) \citep{cirkovic2024emergence} and MRV structures. Incorporating the copulas within the MRV framework, our approach models extremal interlayer dependence with a rigorous theoretical foundation and practical applications.

Applying our UTD estimator to the Reddit interaction network, we uncover how real-world events shape user behavior in online communities:
\begin{enumerate}
    \item Behavoral links in financial subreddits: By analyzing the subreddit r/BitcoinMarkets, we find a closely related pattern between the monthly asset shrinkage ratio of Bitcoins and the UTD of user interactions. Increased market volatility leads to more consistent engagement patterns among high-degree users.
    \item Seasonal impact in sports subreddits: By examining subreddits like r/nba and r/CFB, we demonstrate that the start and end of sports seasons significantly influence user interactions, with engagement patterns reflecting real-world sporting events.
\end{enumerate}
These findings highlight the UTD estimator as a powerful tool for understanding how financial markets and external events drive online behavior. By bridging theoretical network science with practical insights, our study provides valuable implications for researchers and practitioners analyzing complex interaction patterns on digital platforms.

The remainder of this paper is organized as follows. Section \ref{sec:model} introduces the MIRG model and the theoretical background of multivariate regular variation and copula functions. Section \ref{sec: theory} presents the theoretical properties of our proposed UTD estimator. Section \ref{sec:sim} validates our approach through extensive simulation experiments, and Section \ref{sec: DA} demonstrates the practical applicability of our method using the Reddit dataset. Proofs of the theoretical results are provided in the Appendix.

\section{Model}
\label{sec:model}
This section presents the mathematical foundation for the analysis of multilayer network interdependence through the multilayer inhomogeneous random graphs (MIRG). We incorporate multivariate regular variation (MRV) theory and copula functions to characterize the extremal dependence structure of nodal degrees across network layers.

\subsection{Multilayer inhomogeneous random graph}
The MIRG model provides a flexible framework for representing complex multilayer networks. Let $\mathbf{A}(N) = \{A_{ijl}\}_{ijl}$ denote an $N \times N \times L$ adjacency tensor, where $N$ represents the number of nodes and $L$ the number of layers. Each element $A_{ijl} \in \mathbb{N}_0$ indicates the number of edges between nodes $i$ and $j$ in layer $l$. For each layer $l \in [L]$, the submatrix $\{A_{ijl}\}_{ij}$ is symmetric, representing an undirected graph. Also, MIRG allows self-loops and multiple edges between each pair of nodes.

%The model satisfies the following properties:
%\begin{itemize}
%    \item Layer symmetry: For each layer $l \in [L]$, the submatrix $\{A_{ijl}\}_{ij}$ is symmetric, representing undirected graphs;
%    \item Self-loops: Diagonal elements $A_{iil}$ may be non-zero, allowing self-connections;
%    \item Multi-edge support: The model accommodates multiple edges between node pairs.
%\end{itemize}
The connectivity structure is driven by a sequence of independent and identically distributed (i.i.d.) random weight vectors $\mathbf{W}_{[N]} = \{\mathbf{W}_i\}_{i=1}^N$, where each $\mathbf{W}_i = (W_{i1},..., W_{iL}) \in \mathbb{R}_{+}^L$. The component $W_{il}$ represents the connectivity potential of node $i$ in layer $l$, with higher values indicating a greater probability of edge formation. The total connectivity potential in layer $l$ is given by $T_l(N) = \sum_{i=1}^N W_{il}$.

Conditioned on the weight vectors $\mathbf{W}_{[n]}$, the number of edges between nodes $i$ and $j$ in layer $l$ follows a Poisson distribution:
\begin{align}
    \label{def: adj_L_1}
    A_{ijl} \mid \mathbf{W}_{[n]} \stackrel{\text{ind}}{\sim} \operatorname{Poisson}\left(g_l\left(\frac{W_{il}W_{jl}}{T_l(N)}\right)\right), \quad 1 \leq i \leq j \leq N
\end{align}
where $g_l: \mathbb{R}^{+} \rightarrow \mathbb{R}^{+}$ is a layer-specific connection probability function. The properties and selection criteria for $g_l$ are discussed in \cite{cirkovic2024emergence}.
Then by the property of Poisson random variables,
the degree of node $i$ in layer $l$, denoted $D_{il}(N) = \sum_{j=1}^N A_{ijl}$, is distributed as
\begin{align}
    \label{def: degree_L1}
    D_{il}(N) \mid \mathbf{W}_{[N]} \sim \operatorname{Poisson}\left(\sum_{j=1}^N g_l\left(\frac{W_{il}W_{jl}}{T_l(N)}\right)\right), \quad i = 1,...,N.
\end{align}
This formulation establishes the fundamental relationship between nodal weights and degree distributions in the MIRG framework. Details of single-layer and multilayer models can be found in \cite{cirkovic2024emergence, chung2002average, bollobas2007phase}.

The MIRG framework is particularly flexible in capturing interlayer dependence structures, making it well-suited for analyzing complex multilayer networks. Here we specifically focus on modeling interlayer dependence by leveraging tools from multivariate extremes. In the literature, there are two common classes of approaches:
(1) characterizing the support of the limit measure under the 
the \textit{multivariate regular variation} (MRV) framework; and (2)\textit{copula}-based methods. 
We now summarize these two approaches accordingly.
%In the following sections, we will explore these two approaches in detail. Section~\ref{sec: 2.3} introduces the MRV framework and its application to multilayer networks, while Section~\ref{sec:copula} discusses copula-based methods and their role in capturing interlayer dependence structures.

\subsection{Multivariate regular variation}
\label{sec: 2.3}
Multivariate regular variation is a key concept in the extreme value theory to characterize the extremal dependence structure of multivariate extremal events. This section introduces the related definitions essential for our analysis. Further details on the development of MRV can be found in 
\cite{hult:lindskog:2006a, resnick2007heavy, das:mitra:resnick:2013, lindskog:resnick:roy:2014,basrak:planinic:2019,kulik2020heavy,resnickbook:2024}.

Let \(\mathbb{C}_0\) and \(\mathbb{C}\) denote two closed cones in \(\mathbb{R}_{+}^L\), where \(\mathbb{C}_0\) is referred to as the \textit{forbidden zone}. The theoretical foundation of MRV relies on \(\mathbb{M}\)-convergence on the space \(\mathbb{C} \setminus \mathbb{C}_0\).

\begin{definition}[\(\mathbb{M}\)-convergence]
\label{def:convergence}
Let \(\mathbb{M}(\mathbb{C} \setminus \mathbb{C}_0)\) denote the set of Borel measures on \(\mathbb{C} \setminus \mathbb{C}_0\) that are finite on sets bounded away from \(\mathbb{C}_0\). Let \(\mathcal{C}(\mathbb{C} \setminus \mathbb{C}_0)\) represent the set of continuous, bounded, non-negative functions on \(\mathbb{C} \setminus \mathbb{C}_0\) with supports bounded away from \(\mathbb{C}_0\). For \(\mu_n, \mu \in \mathbb{M}(\mathbb{C} \setminus \mathbb{C}_0)\), we say \(\mu_n \rightarrow \mu\) in \(\mathbb{M}(\mathbb{C} \setminus \mathbb{C}_0)\) if
\[
\int f \, d\mu_n \rightarrow \int f \, d\mu
\]
for all \(f \in \mathcal{C}(\mathbb{C} \setminus \mathbb{C}_0)\).
\end{definition}

Using \(\mathbb{M}\)-convergence, the formal definition of multivariate regular variation of distributions for \(\mathbb{C} = \mathbb{R}_{+}^L\) and \(\mathbb{C}_0 = \{\mathbf{0}\}\) is as follows.

\begin{definition}[Multivariate regular variation]
\label{def:2}
A random vector \(\mathbf{Z}\) on \(\mathbb{R}_{+}^L\) (\(L \geq 1\)) is said to have a (standard) regularly varying distribution on \(\mathbb{R}_{+}^L \setminus \{\mathbf{0}\}\) with index \(\alpha > 0\) if there exists a regularly varying scaling function \(b(t)\) with index \(1/\alpha\) and a limit measure \(\nu \in \mathbb{M}(\mathbb{R}_{+}^L \setminus \mathbb{C}_0)\) such that, as \(t \rightarrow \infty\),
\begin{equation}
\label{eq:def2}
t \mathbb{P}\left(\frac{\mathbf{Z}}{b(t)} \in \cdot \right) \rightarrow \nu(\cdot), \quad \text{in } \mathbb{M}(\mathbb{R}_{+}^L \setminus \mathbb{C}_0).
\end{equation}
We denote this by \(\mathbb{P}(\mathbf{Z} \in \cdot) \in \operatorname{MRV}(\alpha, b(t), \nu, \mathbb{R}_{+}^L \setminus \mathbb{C}_0)\).
\end{definition}

Take $L=2$ as an example, and according to \cite{wang20232rv+},
the limit measure \(\nu(\cdot)\) may exhibit distinct forms of asymptotic dependence:
\begin{enumerate}
    \item \textbf{Asymptotic full dependence}: \(\nu\) concentrates on a ray \(\{(x, cx) : x > 0\}\) for some constant \(c > 0\).
    \item \textbf{Asymptotic strong dependence}: \(\nu\) concentrates on a wedge
    \[
    \left\{\boldsymbol{x} \in \mathbb{R}_{+}^2 : a_l x_1 \leq x_2 \leq a_u x_1\right\},
    \]
    where \(0 < a_l < a_u < \infty\).
    \item \textbf{Asymptotic weak dependence}: The support of \(\nu\) covers the entire space \(\mathbb{R}_{+}^2\).
    \item \textbf{Asymptotic independence}: \(\nu((0, \infty)^2) = 0\), indicating that \(\nu\) concentrates solely on the axes.
\end{enumerate}
Such classification provides a comprehensive framework for analyzing extremal dependence in multilayer networks.

\subsection{Copula structure}
\label{sec:copula}
Copula is another common tool to model extremal dependence. It is a multivariate distribution function whose marginals are uniformly distributed on \([0, 1]\). By Sklar's theorem \cite{sklar1959fonctions}, any joint distribution \(F\) of a random vector \(\mathbf{X} = (X_1, X_2, \dots, X_d)\) with marginal distributions \(F_{X_i}(x_i)\) can be expressed as:
\[
F(x_1, x_2, \dots, x_d) = C(F_{X_1}(x_1), F_{X_2}(x_2), \dots, F_{X_d}(x_d)),
\]
where \(C: [0, 1]^d \to [0, 1]\) is the copula function. %This decomposition separates the marginal distributions from the dependence structure, making copulas a powerful tool for modeling extremal dependencies.

The Gumbel copula, part of the Archimedean family, is well-suited for capturing tail dependence, making it ideal for our simulation analysis \cite{embrechts2001modelling}. The Gumbel copula belongs to the family of Archimedean copulas and is widely used for modeling extremal dependence. For a \(d\)-dimensional random vector \(\mathbf{X} = (X_1, X_2, \dots, X_d)\) with marginal distribution functions \(F_{X_i}(x_i)\) for \(i = 1, 2, \dots, d\), the Gumbel copula \(C_{\theta}(u_1, u_2, \dots, u_d)\) is defined as:
\[
C_{\theta}(u_1, u_2, \dots, u_d) = \exp\left(-\left(\sum_{i=1}^d (-\ln u_i)^{\theta}\right)^{1/\theta}\right),
\]
where \(u_i = F_{X_i}(x_i)\) for \(i = 1, 2, \dots, d\), and \(\theta \geq 1\) is the dependence parameter. 

The parameter \(\theta\) indicates the strength of dependence; larger values correspond to stronger tail dependence. When \(\theta = 1\), it reduces to the independence copula \(C(u_1, u_2, \dots, u_d) = \prod_{i=1}^d u_i\), reflecting asymptotic independence where the limit measure \(\nu\) concentrates solely on the axes. As \(\theta \to \infty\), the Gumbel copula approaches the comonotonicity copula, reflecting asymptotic full dependence. Values of \(\theta\) near 1 show asymptotic weak dependence, while larger values indicate stronger dependence.

In the bivariate case (\(d = 2\)), the Gumbel copula simplifies to:
\begin{align}
    C_{\theta}(u, v) = \exp\left(-\left((-\ln u)^{\theta} + (-\ln v)^{\theta}\right)^{1/\theta}\right),
    \label{eq: gumbelcopula_2}
\end{align}
where \(u = F_{X}(x)\) and \(v = F_{Y}(y)\). Other copula families capable of modeling extreme value behavior are discussed in \cite{gudendorf2010extreme}, and our theoretical results derived under the copula framework are presented in Section~\ref{sec: theory}.

\subsection{Upper tail dependence in bilayer networks}
\label{sec:utd}
We now focus on upper tail dependence (UTD) as a practical measure to characterize extremal dependence in bilayer networks.
Within the proposed MIRG framework, the UTD of the weight vector \(\mathbf{W} = (W_1, W_2)\) is a critical quantity for understanding extremal dependence \cite{frahm2005estimating}. It captures the likelihood of simultaneous extremes in the connectivity potentials of the two layers, offering insights into the interdependence of node degrees across layers.

Let \(F_1\) and \(F_2\) denote the cumulative distribution functions (CDFs) of \(W_1\) and \(W_2\), respectively. For a given threshold \(q \in (0, 1)\), define the quantiles \(u_1 = F_1^{-1}(q)\) and \(u_2 = F_2^{-1}(q)\), where \(F_i^{-1}\) represents the inverse CDF. The UTD at level \(q\) is defined as:
\begin{align}
\label{def:utd_W}
\lambda(q) = \mathbb{P}(W_2 > u_2 \mid W_1 > u_1).
\end{align}

The limiting behavior of \(\lambda(q)\) as \(q \to 1\) determines the presence of extremal dependence:
\begin{itemize}
    \item If \(\lim_{q \to 1-} \lambda(q) = 0\), the weights \(W_1\) and \(W_2\) are asymptotically independent.
    \item If \(\lim_{q \to 1-} \lambda(q) > 0\), the weights exhibit upper tail dependence.
\end{itemize}

Denote the upper tail dependence coefficient \(\lambda_U\) as:
\begin{align}
\label{def: utd_W_lim}
\lambda_U := \lim_{q \to 1^-} \lambda(q),
\end{align}
and under the copula framework, \(\lambda_U\) can be expressed as \cite{embrechts2001modelling}:
\begin{equation*}
\lambda_U = \lim_{q \to 1^-} \left( \frac{1-2q+ C(q,q)}{1- q} \right).
\end{equation*}
For the Gumbel copula in particular, the upper tail dependence coefficient \(\lambda_U^{G}\) satisfies:
\begin{equation}
    \lambda_U^{G} = 2 - 2^{1/\theta},
\label{eq: UTD_Gumbel}
\end{equation}
which is a direct link between the copula parameter \(\theta\) and the extremal dependence structure.

%This coefficient quantifies the strength of extremal dependence between layers, with higher values indicating stronger tail dependence. 
Note that the weight vector \(\mathbf{W}\) is typically unobservable in real-world networks; therefore, we extend the concept of upper tail dependence to analyze degree distributions across layers. However, degree data present unique challenges due to their non-i.i.d. nature, arising from complex network topologies and interdependence. Addressing this non-i.i.d. characteristic is essential for accurately capturing tail dependence and understanding interlayer relationships in multilayer networks.

\section{Theoretical results}
\label{sec: theory}
This section establishes the theoretical foundation for estimating and analyzing UTD in multilayer networks. We propose an estimator for the UTD coefficient and prove its consistency under the multilayer inhomogeneous random graph (MIRG) framework.

\begin{theorem}[Consistency of the UTD estimator]
			\label{thm: estimator_UTD}
            Suppose $\alpha>0$, and assume $\mathbb{P}\left(\mathbf{W} \in \cdot\right)$ $\in $ $\operatorname{MRV}$ $\left(\alpha, b(t), \nu, \mathbb{R}_{+}^L \backslash\{\mathbf{0}\}\right)$ for some scaling function $b(t) \in R V_{1 / \alpha}$ and limit measure $\nu$. Consider an $L$-layer MIRG with UTD coefficient between layer $s$ and $m$ \(\lambda_U^{s,m}\) (cf. \eqref{def: utd_W_lim}). Consider the empirical estimator
			\[
			\hat{\lambda}_{D(N)}^{s,m} = \frac{\sum_{i=1}^N \mathbf{1}_{\{D_{is}(N) > \hat{u}_{s,N}\} \cap \{D_{im}(N) > \hat{u}_{m,N}\}}}{\sum_{i=1}^N \mathbf{1}_{\{D_{is}(N) > \hat{u}_{s,N}\}}},
			\]
			where $\hat{u}_{l,N}=\hat{G}_{l,N}^{-1}(1-t_N/N)$ is the $(1-\frac{t_N}{N})$-quantile based on the empirical distribution function $\hat{G}_{l,N}(x)=\frac{1}{N}\sum_{i = 1}^N\mathbf{1}_{\{D_{il}(N)\leq x\}}$, $l = s, m$. Then we have 
			\[
			\hat{\lambda}_{D(N)}^{s,m}  \stackrel{p}{\longrightarrow} \lambda_U^{s,m},
			\]
            as \(N \to \infty\).
		\end{theorem}
        
        The proof of Theorem~\ref{thm: estimator_UTD} is provided in \ref{app1}. The consistency of $\hat{\lambda}_{D(N)}^{s,m}$ justifies the way to quantify extremal dependence in large-scale multilayer networks. An immediate consequence of Theorem~\ref{thm: estimator_UTD} is
        the special case where the dependence structure is governed by a copula function.

\begin{corollary}\label{cor: general_copula}
	Assume the weight vector $\mathbf{W} = (W_1, \ldots, W_L)$ follows a distribution with pairwise copulas $C_{s,m}(u,v)$ for layers $s$ and $m$, and marginal distributions that are regularly varying with index $\alpha > 0$. Under the MRV assumptions of Theorem~\ref{thm: estimator_UTD}, the pairwise upper tail dependence coefficient satisfies:
	\[
	\hat{\lambda}_{D(N)}^{s,m} \stackrel{p}{\longrightarrow} \lambda_{U}^{s,m} = \lim_{q \to 1^-} \left( \frac{1-2q+ C_{s,m}(q,q)}{1- q} \right).
	\]
    Specifically, consider the Gumbel copula \eqref{eq: gumbelcopula_2} with $\theta_{s,m} \geq 1$ for layers $s$ and $m$, we obtain:
\[
\hat{\lambda}_{D(N)}^{s,m} \stackrel{p}{\longrightarrow} \lambda_{U}^{G, s,m} = 2 - 2^{1/\theta_{s,m}}.
\]
\end{corollary}

Corollary~\ref{cor: general_copula} demonstrates that the pairwise UTD coefficient is explicitly determined by the copula $C_{s,m}(u,v)$, even in the presence of the nonlinear network formation mechanism. This result highlights the robustness of the MRV framework in capturing pairwise extremal dependence structures, providing a powerful tool for analyzing multilayer networks with complex dependency patterns across different layers.

\section{Simulation study}
\label{sec:sim}
We now give a comprehensive simulation study to investigate the relationship between the tail dependence of degree distributions in bilayer networks (i.e. $L=2$) and that of weight vectors \(\mathbf{W}\) under various dependence scenarios. 
We follow the procedure given in Algorithm~\ref{alg:simulation} to generate $n=1000$ realizations of a specific MIRG with \(g_l(x) = x \), which corresponds to the Norros-Reittu model \cite{norros2006conditionally}. The algorithm consists of three main steps: weight vector generation, network construction, and degree calculation with UTD estimation. %Each step is designed to ensure a robust exploration of the dependence structures in multilayer networks.

For $L=2$, we simplify the notation by omitting the layer indices, using \(\hat{\lambda}(q)\) and \(\hat{\lambda}_{D(N)}\) to denote the UTD estimators for the weight vectors and the degree distributions, respectively.

\begin{algorithm}[H]
    \caption{Simulation procedure for investigating tail dependence in bilayer networks}
    \label{alg:simulation}
    \begin{algorithmic}[1]
        \Require Number of networks $n = 1000$, dependence structures (Gumbel copula or a particular characterization of $\nu$)
        \Statex \hspace*{-\algorithmicindent}\textbf{Output:} UTD estimates for weight vectors $\hat{\lambda}(q)$ and degree distributions $\hat{\lambda}_{D(N)}$
        \Statex
        \For{$i = 1$ to $n$}
            \State Generate weight vectors $\mathbf{W}_{[N]}$ using Gumbel copula or following a specific characterization of $\nu$.
            \State Construct bilayer network using MIRG and compute adjacency matrices.
            \State Calculate node degrees and estimate UTD using \textit{taildep} from the R package \textit{extRemes}.
        \EndFor
    \end{algorithmic}
\end{algorithm}

\subsection{Weight vector generation based on Gumbel copula}
\begin{figure}[h]
	\centering
	\begin{subfigure}{0.45\textwidth}
		\centering
		\includegraphics[width=\linewidth]{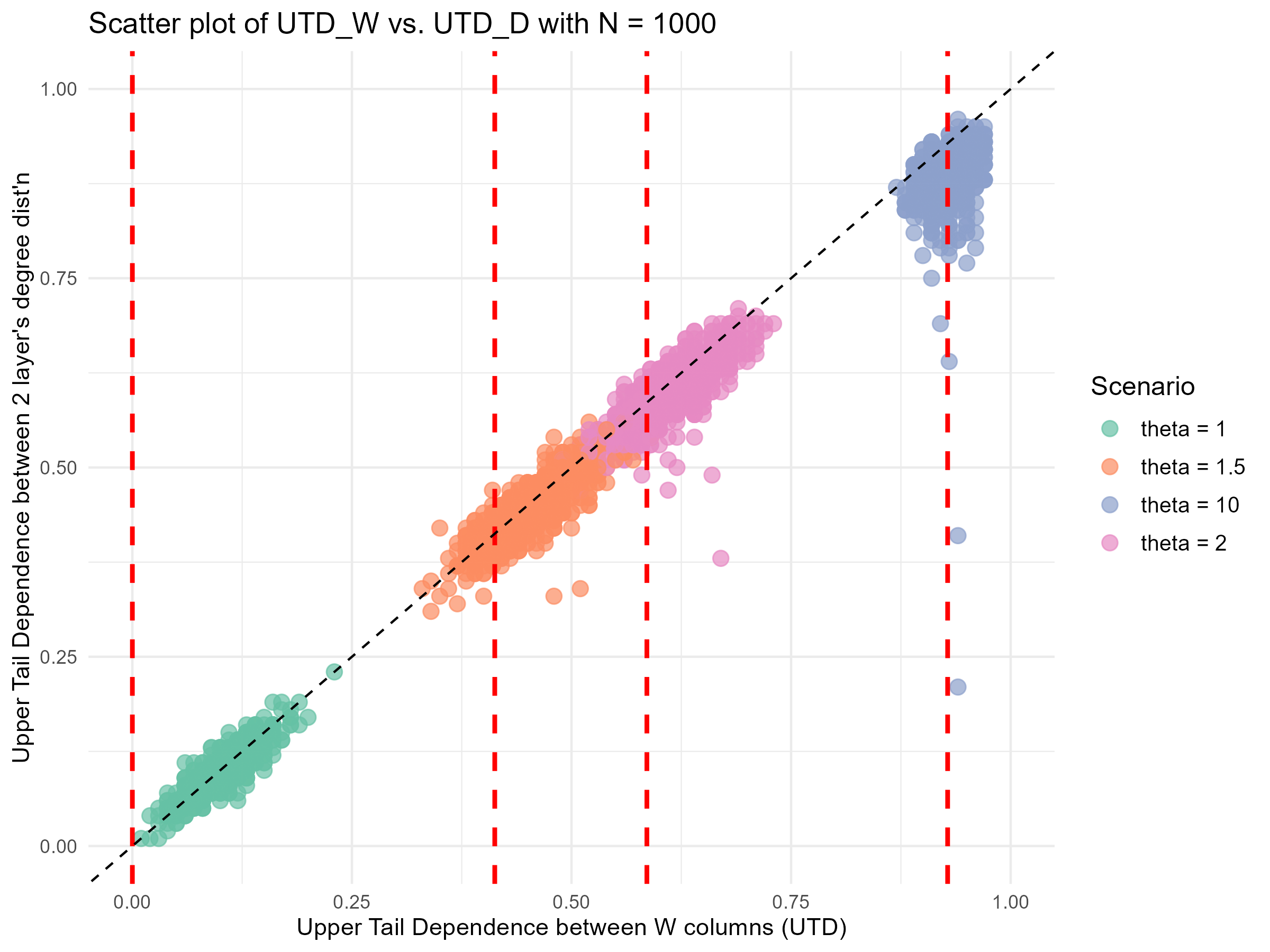}
		\caption{Case when $N = 1000$}
		\label{fig:copula_1k}
	\end{subfigure}
	\hfill
	\begin{subfigure}{0.45\textwidth}
		\centering
		\includegraphics[width=\linewidth]{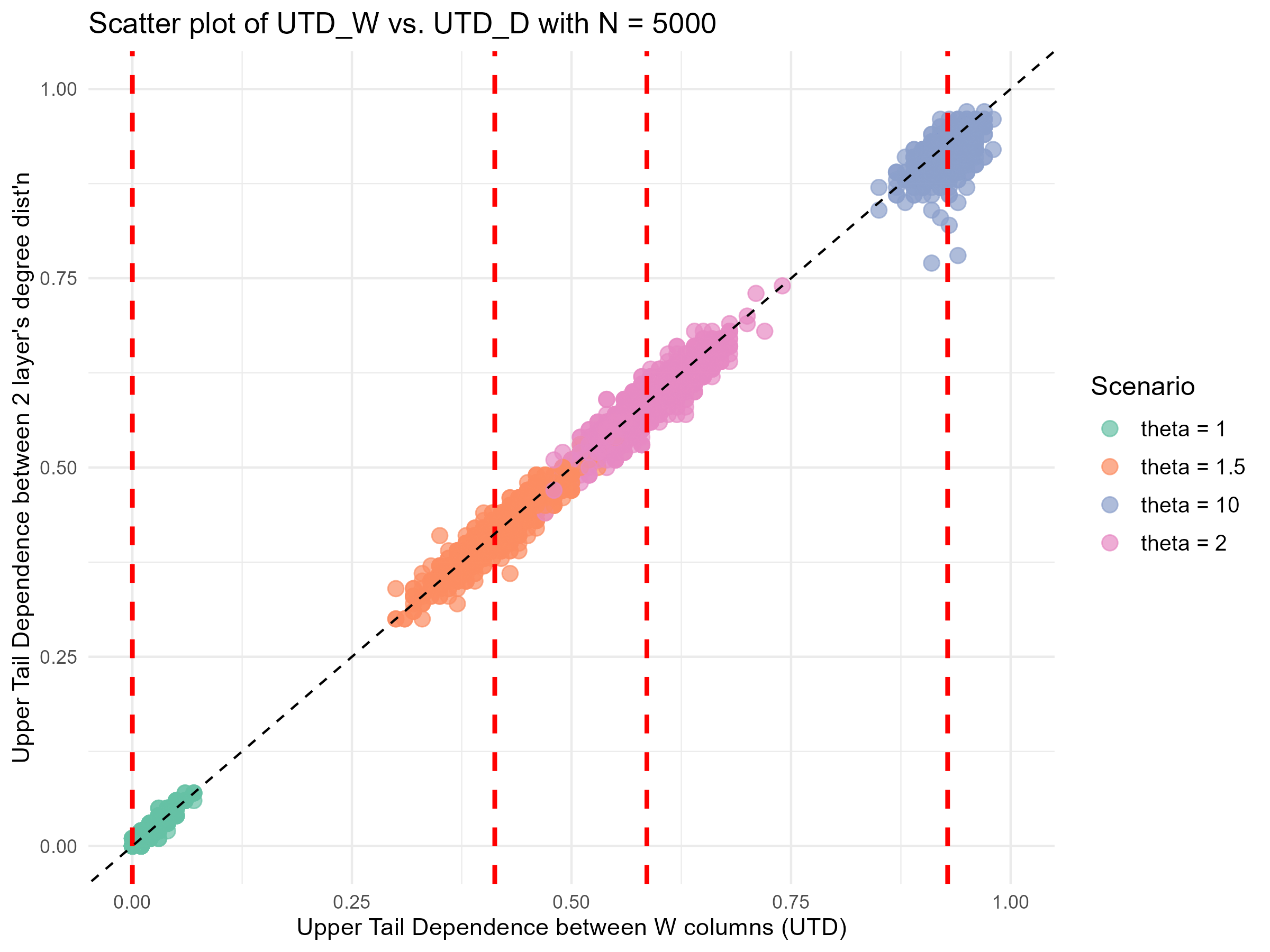}
		\caption{Case when $N = 5000$}
		\label{fig:copula_5k}
	\end{subfigure}
	
	\begin{subfigure}{0.45\textwidth}
		\centering
		\includegraphics[width=\linewidth]{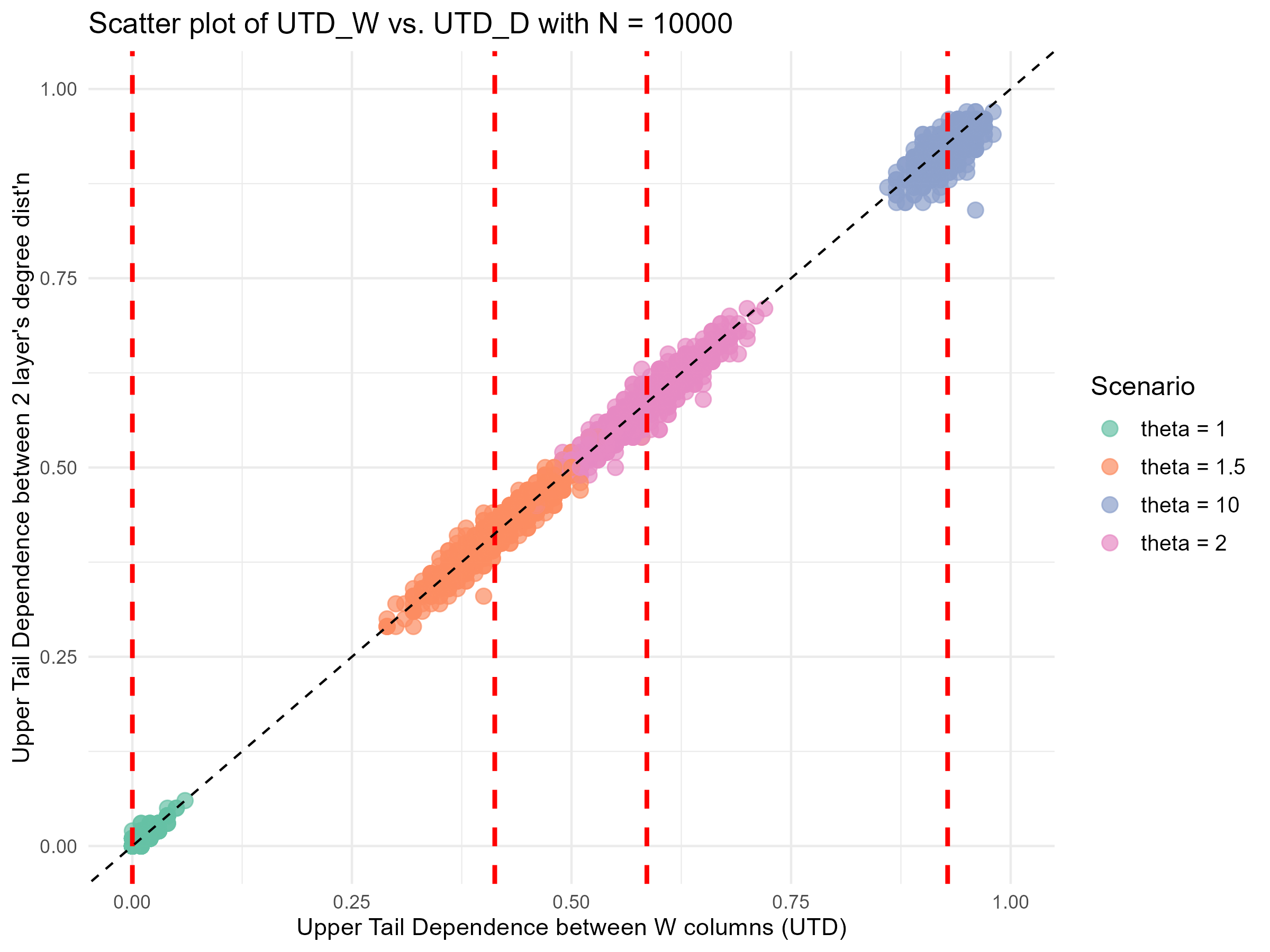}
		\caption{Case when $N = 10000$}
		\label{fig:copula_1W}
	\end{subfigure}
	\hfill
	\begin{subfigure}{0.45\textwidth}
		\centering
		\includegraphics[width=\linewidth]{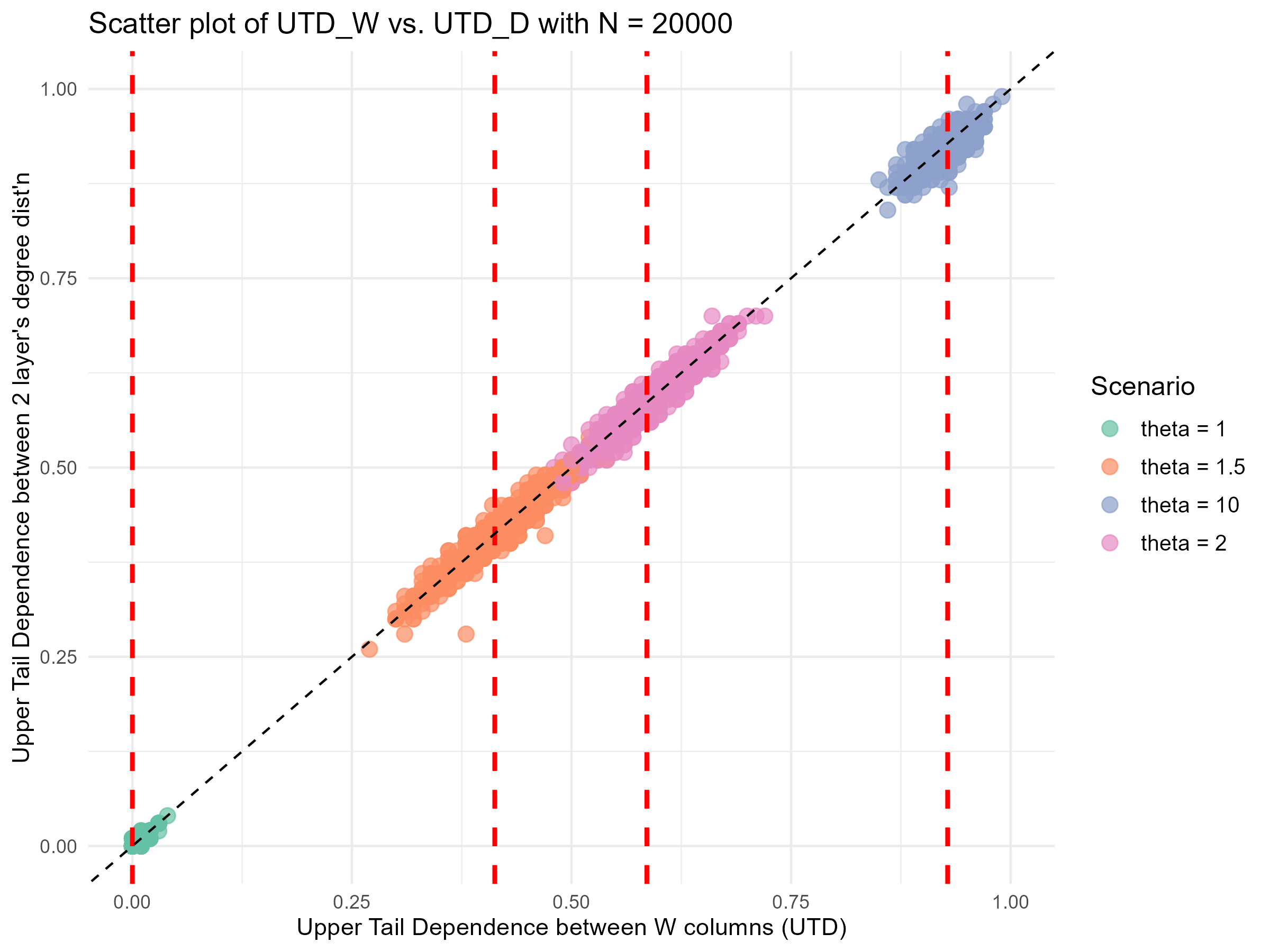}
		\caption{Case when $N = 20000$}
		\label{fig:copula_2W}
	\end{subfigure}
	\caption{This scatterplot shows the relationship between $\hat{\lambda}(q)$ and $\hat{\lambda}_{D(N)}$ for sample sizes $N = 1000, 5000, 10000,$ and $20000$, with $\mathbf{W}$ generated by the Gumbel copula. The red dashed line indicates the true UTD value for each case. Detailed results, including average UTD and MSE, are presented in Table~\ref{tab: sim_copula} and Figure~\ref{fig: copula_mse}.}
	\label{fig: simu_copula_all}
\end{figure}
We start by using a Gumbel copula to generate $\mathbf{W}$. The relationship between the copula parameter \(\theta\) and the UTD coefficient \(\lambda_U^G\) is given by \eqref{eq: UTD_Gumbel} which allows precise control over the UTD by adjusting \(\theta\).
We consider \(\theta = 1, 1.5, 2, 10\), corresponding to true UTD values \(\lambda_U^G = 0, 0.4126, 0.5858, 0.9282\). For each \(\theta\), we generate weight vectors by sampling \(n = 1000\) samples from a Gumbel copula with the specified \(\theta\). These samples are then transformed using the inverse quantile function of a Pareto distribution with parameters \(\alpha = 1.1\) and \(k = 20\) to ensure the heavy-tailed marginal distributions.

The UTD is then estimated for both the weight vectors and the degree sequences using the \textit{taildep} function. We consider four network sizes: \(N = 1000, 5000, 10000, 20000\), with thresholds \(q = 0.9, 0.98, 0.99, 0.995\) corresponding to the top 100 nodes in each case.

\begin{figure}[ht]
	\centering
	\includegraphics[width = \textwidth]{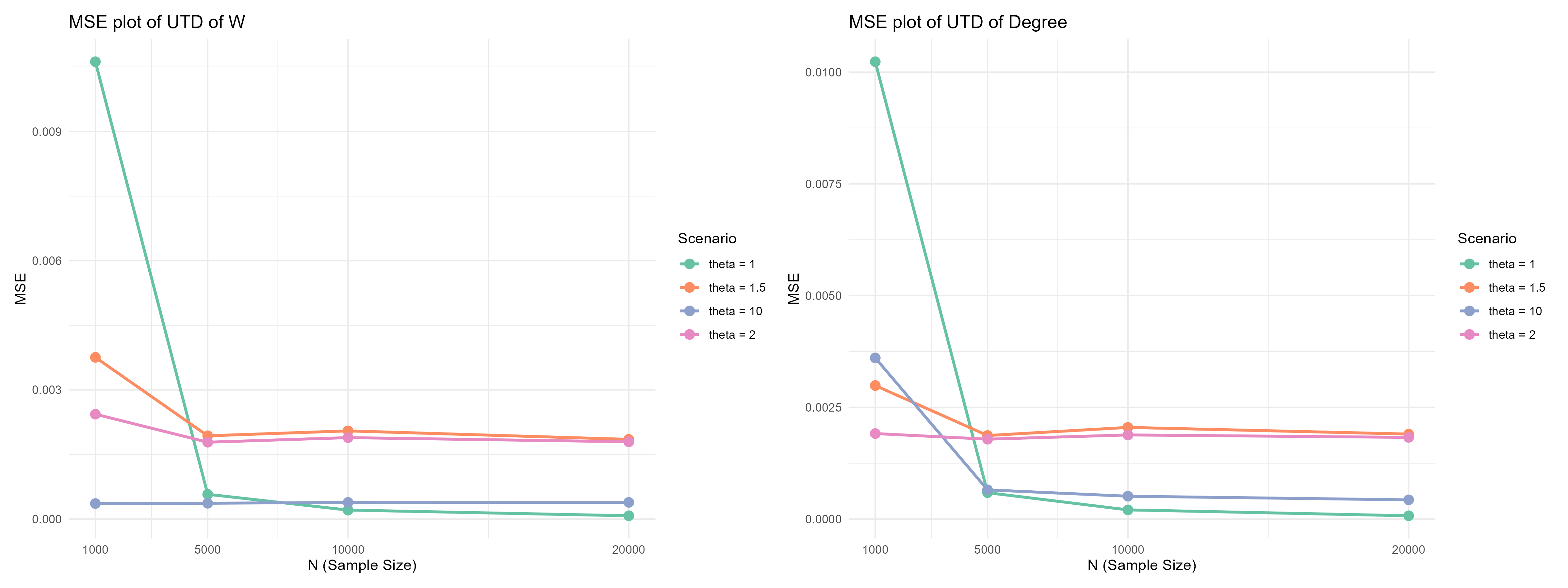}
	\caption{Mean squared error (MSE) plots for the estimated UTD values are shown. The left panel corresponds to the UTD of $\mathbf{W}$, while the right panel corresponds to the UTD of the degree distribution. Theoretical UTD values for $\mathbf{W}$ are based on the Gumbel copula. The plots display MSEs for different values of $\theta$ ($\theta = 1, 1.5, 2, 10$) across varying sample sizes $N$. The x-axis denotes the sample size $N$, and the y-axis denotes the MSE.}
	\label{fig: copula_mse}
\end{figure}

The scatterplots in Figure~\ref{fig: simu_copula_all} show the relationship between \(\hat{\lambda}(q)\) and \(\hat{\lambda}_{D(N)}\) for different network sizes. The sample points are concentrated around the line \(y = x\), indicating strong agreement between the estimated UTD values. The average UTD values and mean squared errors (MSEs) are presented in Table~\ref{tab: sim_copula} and Figure~\ref{fig: copula_mse}, respectively. The results demonstrate that the MSE decreases as the sample size \(N\) increases, confirming the consistency of the estimator.
\begin{table}[h]
	\centering
	\resizebox{\textwidth}{!}{
		\begin{tabular}{|c|c|c|c|c|c|c|c|c|c|}
			\hline
			\multicolumn{2}{|c|}{Gumbel copula} & \multicolumn{2}{c|}{N = 1000} & \multicolumn{2}{c|}{N = 5000} & \multicolumn{2}{c|}{N = 10000} & \multicolumn{2}{c|}{N = 20000} \\
			\cline{3-10}
			\multicolumn{2}{|c|}{parameter} & 	$\hat{\lambda}(q)$ & $\hat{\lambda}_{D(N)}$ & $\hat{\lambda}(q)$ & $\hat{\lambda}_{D(N)}$ & $\hat{\lambda}(q)$ & $\hat{\lambda}_{D(N)}$ & $\hat{\lambda}(q)$ & $\hat{\lambda}_{D(N)}$  \\
			\hline
			$\theta = 1$ & $\lambda_U^G = 0$& 0.0990 & 0.0973 & 0.0199 & 0.0199 & 0.0104 & 0.0103  & 0.0050  & 0.0050  \\
			\hline
			$\theta = 1.5$ & $\lambda_U^G = 0.4126$ & 0.4575  & 0.4486 & 0.4205 & 0.4179 & 0.4161 & 0.4151  & 0.4134  & 0.4128  \\
			\hline
			$\theta = 2$ & $\lambda_U^G = 0.5858$ & 0.6147 & 0.6012 & 0.5911 & 0.5872 &  0.5885 & 0.5870  & 0.5870  & 0.5859  \\
			\hline
			$\theta = 10$ & $\lambda_U^G = 0.9282$ & 0.9304 & 0.8832 & 0.9288 & 0.9268 & 0.9267 &  0.9189  & 0.9265  & 0.9221  \\
			\hline
		\end{tabular}}
	\caption{This table provides simulation results for the UTD means of $\mathbf{W}$ generated by the Gumbel copula across different sample sizes $N$. Each value is the average of 1000 estimates for $\hat{\lambda}(q)$ and $\hat{\lambda}_{D(N)}$. The table includes the Gumbel copula parameter $\theta$ and the true UTD value $\lambda_U$, with results shown for $N = 1000$, $5000$, $10000$, and $20000$.}
	\label{tab: sim_copula}
\end{table}

\subsection{Weight vector generation based on multivariate regular variation}
Next, we consider generating weight vectors \(\mathbf{W}\) following the 4-case classification. Suppose \(\mathbf{W} = (W_1, W_2) = (V\Theta, V(1-\Theta))\), where \(V \sim \text{Pareto}(\alpha, k)\) with \(\alpha = 1.1\) and \(k = 20\), and \(\Theta\) is independent of \(V\). We examine four scenarios for \(\Theta\) to model different dependence structures:
\begin{itemize}
	\item \textbf{Asymptotic full dependence}: \(\Theta = 0.5\) (perfect dependence);
	\item \textbf{Asymptotic strong dependence}: \(\Theta \sim \text{Beta}(0.1, 0.1, 0.4, 0.6)\). Here, $Y \sim \operatorname{Beta}\left(b_1, b_2, c_1, c_2\right)$ if $Y=\left(c_2-c_1\right) X+c_1$ for $X \sim \operatorname{Beta}\left(b_1, b_2\right)$ and $b_1, b_2>0$, $c_2>c_1 \geq 0$;
	\item \textbf{Asymptotic weak dependence}: \(\Theta \sim \text{Beta}(0.5, 0.5)\);
	\item  \textbf{Asymptotic independence}: \(\Theta \sim \text{Bernoulli}(0.5)\).
\end{itemize}
For each scenario, we generate networks with \(N = 1000, 5000, 10000, 20000\) nodes and compute the UTD using the same threshold selection method as in the Gumbel copula case.

\begin{figure}[h]
	\centering
	\begin{subfigure}{0.45\textwidth}
		\centering
		\includegraphics[width=\linewidth]{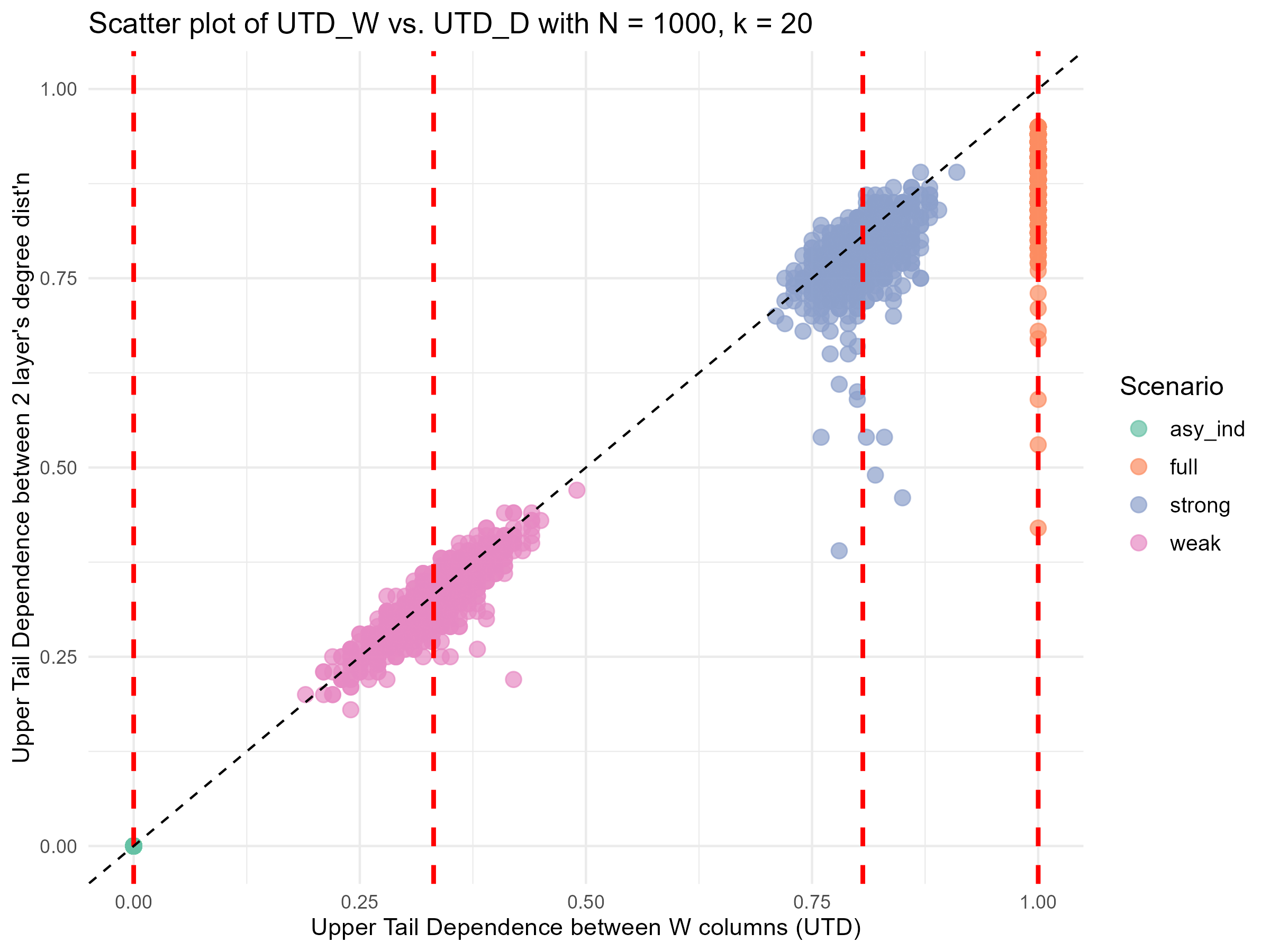}
		\caption{Case when $N = 1000$}
		\label{fig:1k}
	\end{subfigure}
	\hfill
	\begin{subfigure}{0.45\textwidth}
		\centering
		\includegraphics[width=\linewidth]{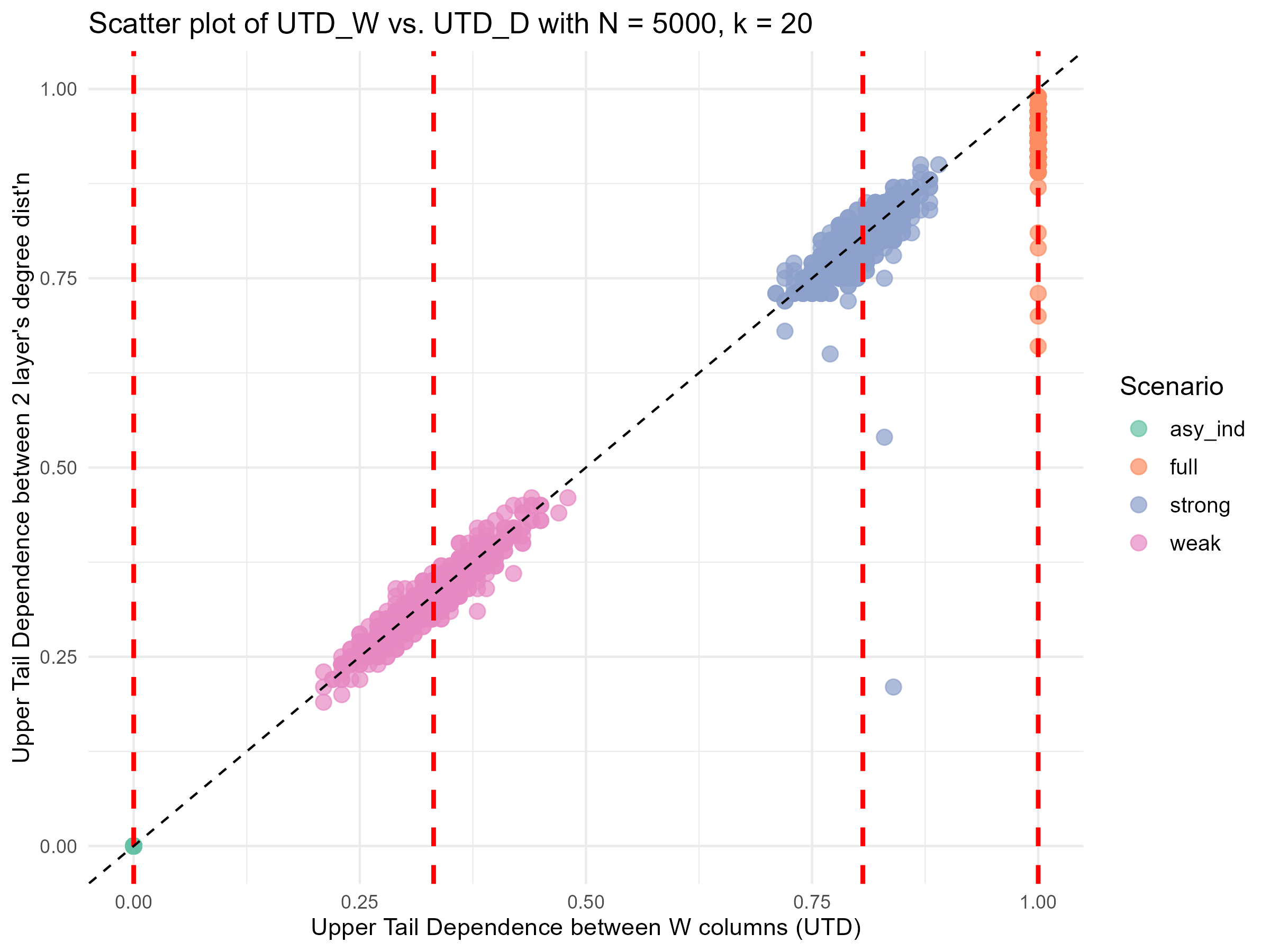}
		\caption{Case when $N = 5000$}
		\label{fig:5k}
	\end{subfigure}
	
	\begin{subfigure}{0.45\textwidth}
		\centering
		\includegraphics[width=\linewidth]{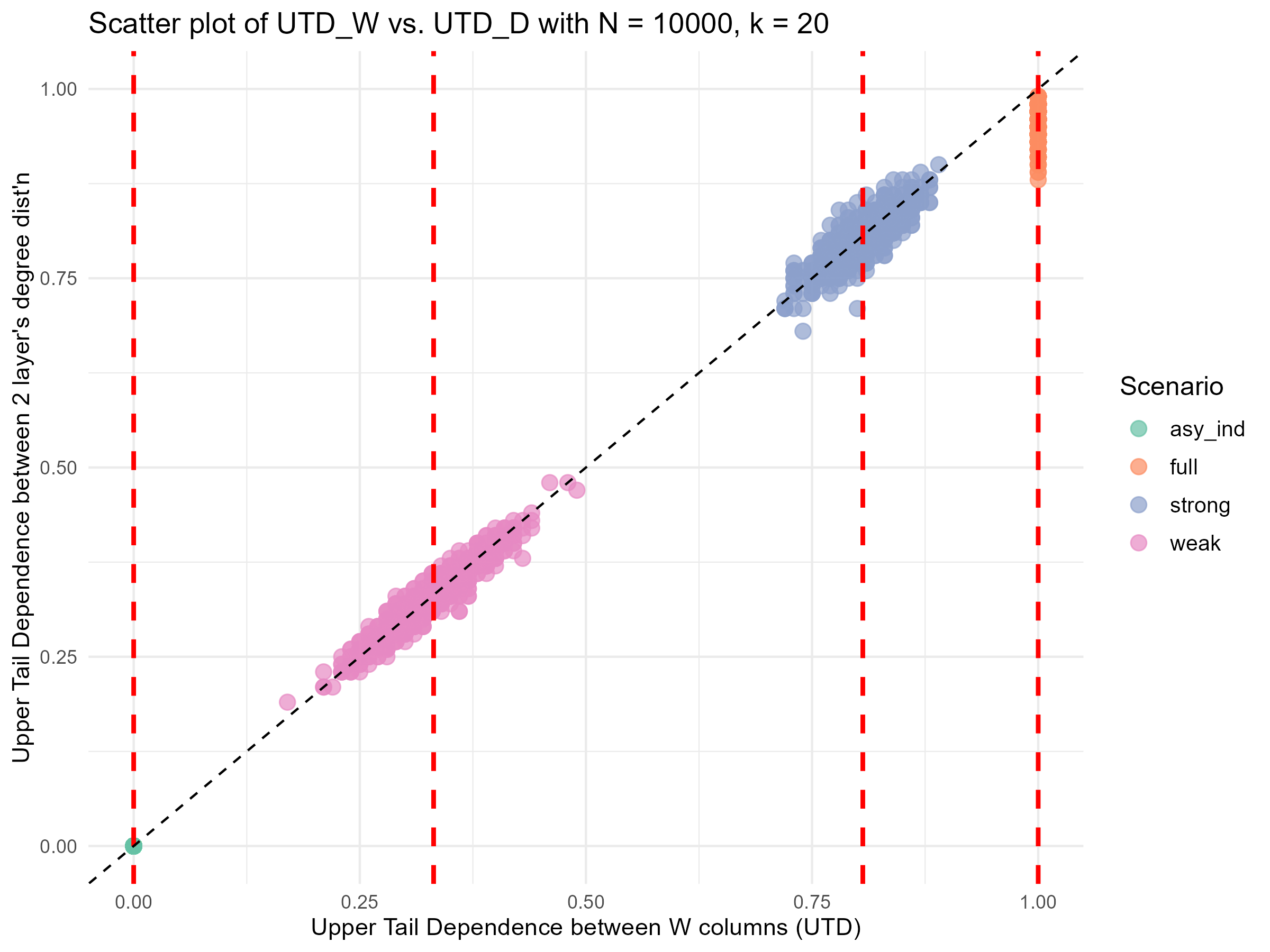}
		\caption{Case when $N = 10000$}
		\label{fig:1W}
	\end{subfigure}
	\hfill
	\begin{subfigure}{0.45\textwidth}
		\centering
		\includegraphics[width=\linewidth]{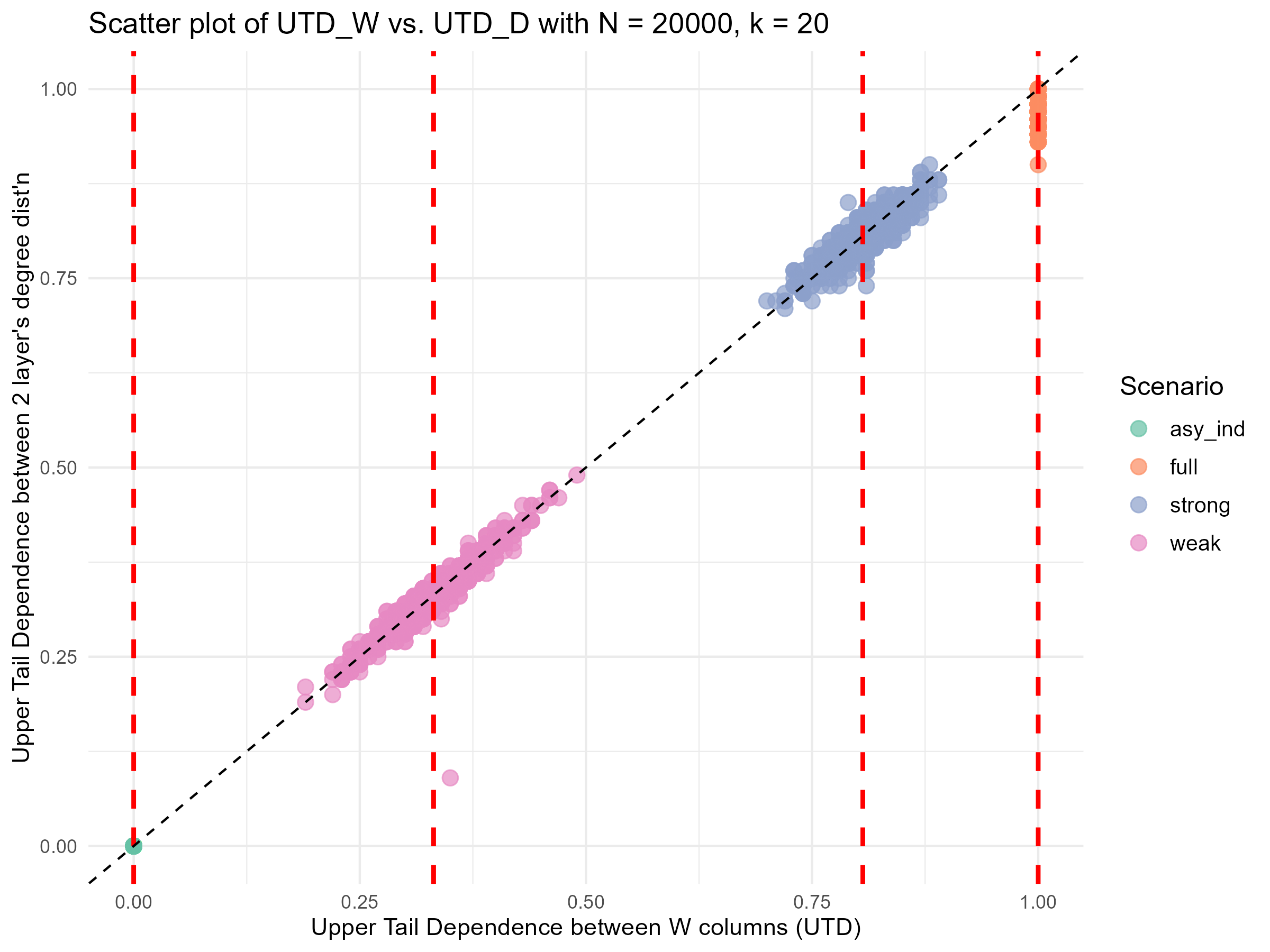}
		\caption{Case when $N = 20000$}
		\label{fig:2W}
	\end{subfigure}
	\caption{This scatterplot shows the relationship between $\hat{\lambda}(q)$ and $\hat{\lambda}_{D(N)}$ for sample sizes $N = 1000$, $5000$, $10000$, and $20000$, with $\mathbf{W}$ generated by MRV. The red dashed line indicates the true UTD value for each case. Detailed results, including average UTD and MSE, are provided in Table~\ref{tab: simu} and Figure~\ref{fig: mrv_mse}.}
	\label{fig: simu_all}
\end{figure}

The scatterplots in Figure~\ref{fig: simu_all} show the relationship between \(\hat{\lambda}(q)\) and \(\hat{\lambda}_{D(N)}\) for different dependence structures and network sizes. The results in Table~\ref{tab: simu} demonstrate that the estimators \(\hat{\lambda}(q)\) and \(\hat{\lambda}_{D(N)}\) effectively distinguish between asymptotic independence, weak dependence, strong dependence, and full dependence.
\begin{table}[h]
	\centering
		\resizebox{\textwidth}{!}{
	\begin{tabular}{|c|c|c|c|c|c|c|c|c|c|}
		\hline
		\multicolumn{2}{|c|}{Dependence}& \multicolumn{2}{c|}{N = 1000} & \multicolumn{2}{c|}{N = 5000} & \multicolumn{2}{c|}{N = 10000} & \multicolumn{2}{c|}{N = 20000} \\
		\cline{3-10}
		\multicolumn{2}{|c|}{Structure} & $\hat{\lambda}(q)$ & $\hat{\lambda}_{D(N)}$ & $\hat{\lambda}(q)$ & $\hat{\lambda}_{D(N)}$ & $\hat{\lambda}(q)$ & $\hat{\lambda}_{D(N)}$ & $\hat{\lambda}(q)$ & $\hat{\lambda}_{D(N)}$  \\
		\hline
		Asy. indep. &\(\lambda_U = 0\)  & 0 & 0 & 0 & 0 & 0  & 0  & 0  & 0  \\
		\hline
		Weak asy. dep. &\(\lambda_U = 0.3316\)  & 0.3300 & 0.3212 & 0.3292 & 0.3272 &  0.3302 & 0.3288  & 0.3286  & 0.3279  \\
            \hline
		Strong asy. dep. &\(\lambda_U = 0.8061\)  & 0.8057 & 0.7811 &0.8062 & 0.8018 & 0.8071 & 0.8050  & 0.8060  & 0.8046  \\
		\hline
		Full asy. dep. &\(\lambda_U = 1\)  & 1 & 0.8873 & 1 & 0.9567 & 1 &  0.9613  & 1  & 0.9715  \\
		\hline
	\end{tabular}
}
	\caption{This table presents the simulation results of the UTD means for $\mathbf{W}$ generated by MRV under different sample sizes $N$. Results are based on 1000 repeated experiments. Each value represents the average of 1000 estimates of $\hat{\lambda}(q)$ (computed from sample $\mathbf{W}_{[N]}$) or $\hat{\lambda}_{D(N)}$ (computed from degree sample sequences). Results for different sample sizes $N = 1000$, $5000$, $10000$, and $20000$ are shown in separate columns. The true UTD values $\lambda_U$ are obtained through numerical methods, serving as a benchmark for evaluating the accuracy of the estimated values derived from the simulations and different sampling approaches.}
	\label{tab: simu}
\end{table}

\begin{figure}[h]
	\centering
	\includegraphics[width=\textwidth]{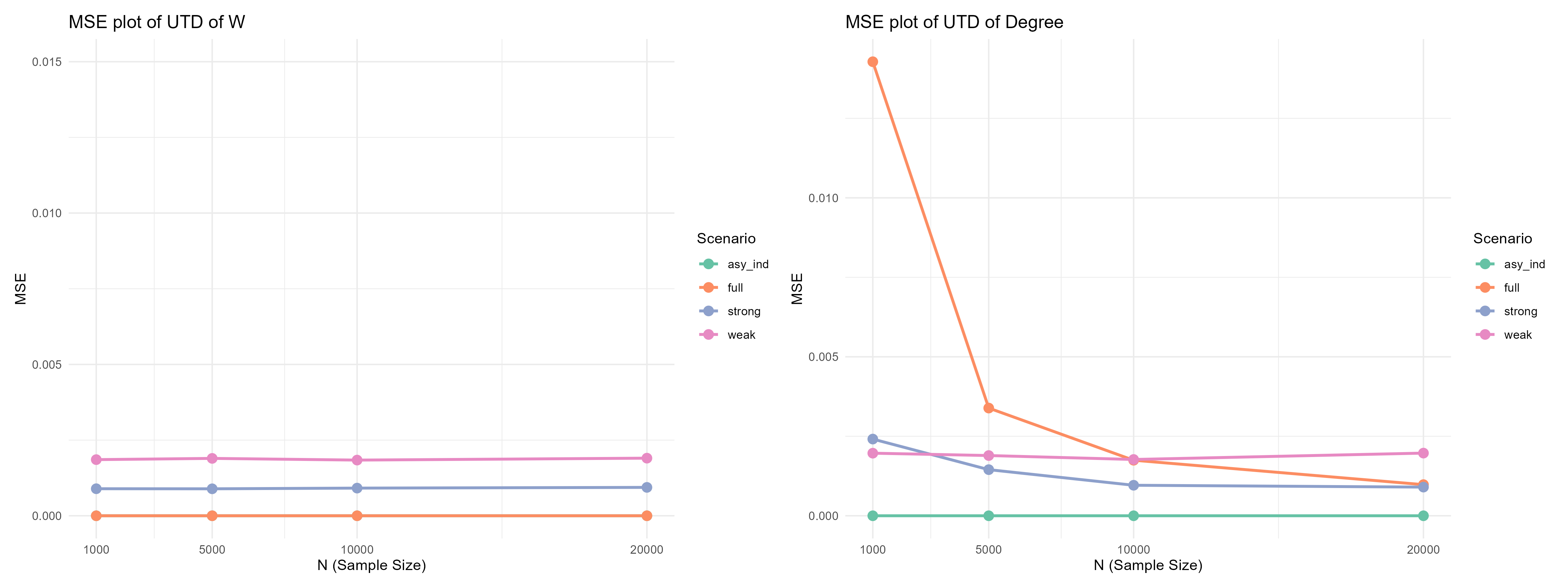}
	\caption{Mean squared error (MSE) plots for the estimated UTD values are shown. The left panel corresponds to the UTD of $\mathbf{W}$, which shows that under small sample conditions, the estimation accuracy has reached near the theoretical limit. The right panel corresponds to the UTD of the degree distribution. The x-axis represents the sample size $N$, and the y-axis represents the MSE. Each line with its corresponding points shows the MSE values for different scenarios across varying sample sizes.}
	\label{fig: mrv_mse}
\end{figure}

 As the sample size \(N\) increases, the estimates become more accurate, with \(\hat{\lambda}_{D(N)}\) converging to the true UTD values \(\lambda_U\). The MSE plot in Figure~\ref{fig: mrv_mse} further confirms the consistency of the estimator, showing a clear decrease in MSE as \(N\) increases.

The simulation results highlight the effectiveness of the proposed UTD estimator in capturing extremal dependence structures in multilayer networks. Both the Gumbel copula and MRV frameworks demonstrate that the estimator is consistent and robust across different dependence scenarios and network sizes. These findings provide strong empirical support for the theoretical results presented in Section~\ref{sec: theory}.

\begin{table}[h]
        \centering
        \begin{tabular}{|c|c|c|}
            \hline
            scenario & $t_N$Var($\hat{\lambda}(q)$) & $t_N$Var($\hat{\lambda}_{D(N)}$) \\
            \hline
            $\theta = 1$   & 0.0050651 & 0.0049647 \\
            \hline
            $\theta = 1.5$ & 0.1851259 & 0.1903595 \\
            \hline
            $\theta = 2$   & 0.1794354 & 0.1829655 \\
            \hline
            $\theta = 10$  & 0.0384474 & 0.0392290 \\
            \hline
        \end{tabular}
        \caption{Asymptotic variances of $\hat{\lambda}_{D(N)}$ and $\hat{\lambda}(q)$ under different scenarios with $N = 20000$ under the copula framework}
        \label{tab:variances_copula}
    \end{table}
    \begin{table}[h]
        \centering
        \begin{tabular}{|c|c|c|}
            \hline
            scenario & $t_N$Var($\hat{\lambda}(q)$) & $t_N$Var($\hat{\lambda}_{D(N)}$) \\
            \hline
            Strong asymptotic dep.    & 0.0938457 & 0.0900171 \\
            \hline
            Weak asymptotic dep.     & 0.1896354 &  0.1958807 \\
            \hline
        \end{tabular}
        \caption{Asymptotic variances of $\hat{\lambda}_{D(N)}$ and $\hat{\lambda}(q)$ under different scenarios with $N = 20000$ under the MRV framework. For the cases of asymptotic independence and full asymptotic dependence, since the upper tail dependence of $\mathbf{W}$ is strictly equal to 0 or 1 respectively, the variances are 0. Thus, these two special cases are not presented in the table. }
        \label{tab:variances_mrv}
    \end{table}

%\tw{Need modifications on wording. We haven't proved the normality; it's just sim evidence suggesting it might be asy normal with the same asy var.}
%As a direct corollary of Theorem 1 in \cite{cirkovic2024tail}, the variances of the two estimators \(\hat{\lambda}_{D(N)}\) and \(\hat{\lambda}(q)\) are equal. 
To further explore the asymptotic distribution of $\hat{\lambda}_
{D(N)}$, simulation results in Tables~\ref{tab:variances_copula} and \ref{tab:variances_mrv} indicate that, for different values of UTD, the asymptotic variances of \(\hat{\lambda}_{D(N)}\) and \(\hat{\lambda}(q)\) are approximately equal. This aligns with the findings in \cite[Theorem 1]{cirkovic2024tail} for single-layer Norros-Reittu graphs. Furthermore, the UTD estimates also pass the Shapiro-Wilk normality test, suggesting that the sequence may exhibit asymptotic normality. We leave the formal justification of the asymptotic normality of the UTD estimator in the multilayer setup for future work.
%This observation, combined with the consistency of the variance estimates, reinforces the robustness of our findings and the suitability of the assumptions underlying the theoretical framework.

\section{Uncovering real-world influences on online behavior with UTD}
\label{sec: DA}
We analyze a dataset of posts and comments from \textit{reddit.com}, a popular platform where users form topic-based discussion communities called subreddits. The dataset consists of monthly user interaction networks from the year 2014, covering 2,046 subreddits \citep{hamilton2017loyalty}. Two types of networks are included: (1) chain-based interaction networks, where users comment within linear chains with at most two intervening comments, and (2) reply-based interaction networks, where connections are formed only when a user directly replies to another. The original networks are directed, with links pointing from the replier to the user being replied to.

\subsection{Reddit interaction data overview and preprocessing}
The dataset includes approximately \(10^8\) comments from \(10^7\) users across \(10^4\) communities in 2014. We focus on three subreddits: “\textit{nba}” (discussions on NBA games, teams, and players), “\textit{CFB}” (college football discussions), and “\textit{BitcoinMarkets}” (cryptocurrency trading and market trends). For “\textit{nba}” and “\textit{CFB}”, we use chain-based interaction networks, while for “\textit{BitcoinMarkets}”, we use reply-based interaction networks. For each subreddit, we construct 11 monthly interaction networks (aligned with ISO 4-week periods from January 27 to November 30, 2014), represented as directed adjacency lists. December and January data are excluded due to data quality issues.

We convert the networks to undirected graphs by ignoring reply directions. For each month, we identify the intersection of nodes across the two networks using user IDs. Degrees are calculated based on undirected connections: mutual replies are counted as two edges (incrementing both nodes' degrees by 2), whereas one-way replies are counted as one edge (incrementing both nodes' degrees by 1). Only edges between nodes in the intersection set are considered. Furthermore, we have verified that the tail index of the data from the three subreddits is between 1 and 2, aligning with the assumptions of the MIRG framework. This validates its applicability to our real-world data analysis. The UTD of the degree distribution is computed using the \textit{taildep} function from the \textit{extRemes} package.

\subsection{UTD and the influence of market trends on financial community engagement}
We explore the relationship between the monthly asset shrinkage ratio of Bitcoins and the UTD of the user reply network in the \textit{BitcoinMarkets} subreddit. This subreddit serves as a hub where Bitcoin enthusiasts, investors, and traders discuss cryptocurrency-related topics.

We calculate UTD for consecutive months, which measures the consistency of reply frequencies among high-degree users. A high UTD indicates a strong correlation in user engagement at the upper end of the reply frequency distribution. We compute the monthly asset shrinkage ratio as:
\[
\frac{\text{Initial Price} - \text{Final Price}}{\text{Initial Price}},
\]
where the initial and final prices are the closing prices on the first and last days of each 4-week period, respectively. This ratio represents the negative monthly return, capturing Bitcoin’s relative decline over the period.

\begin{figure}[h]
	\centering
	\includegraphics[width=0.9\textwidth]{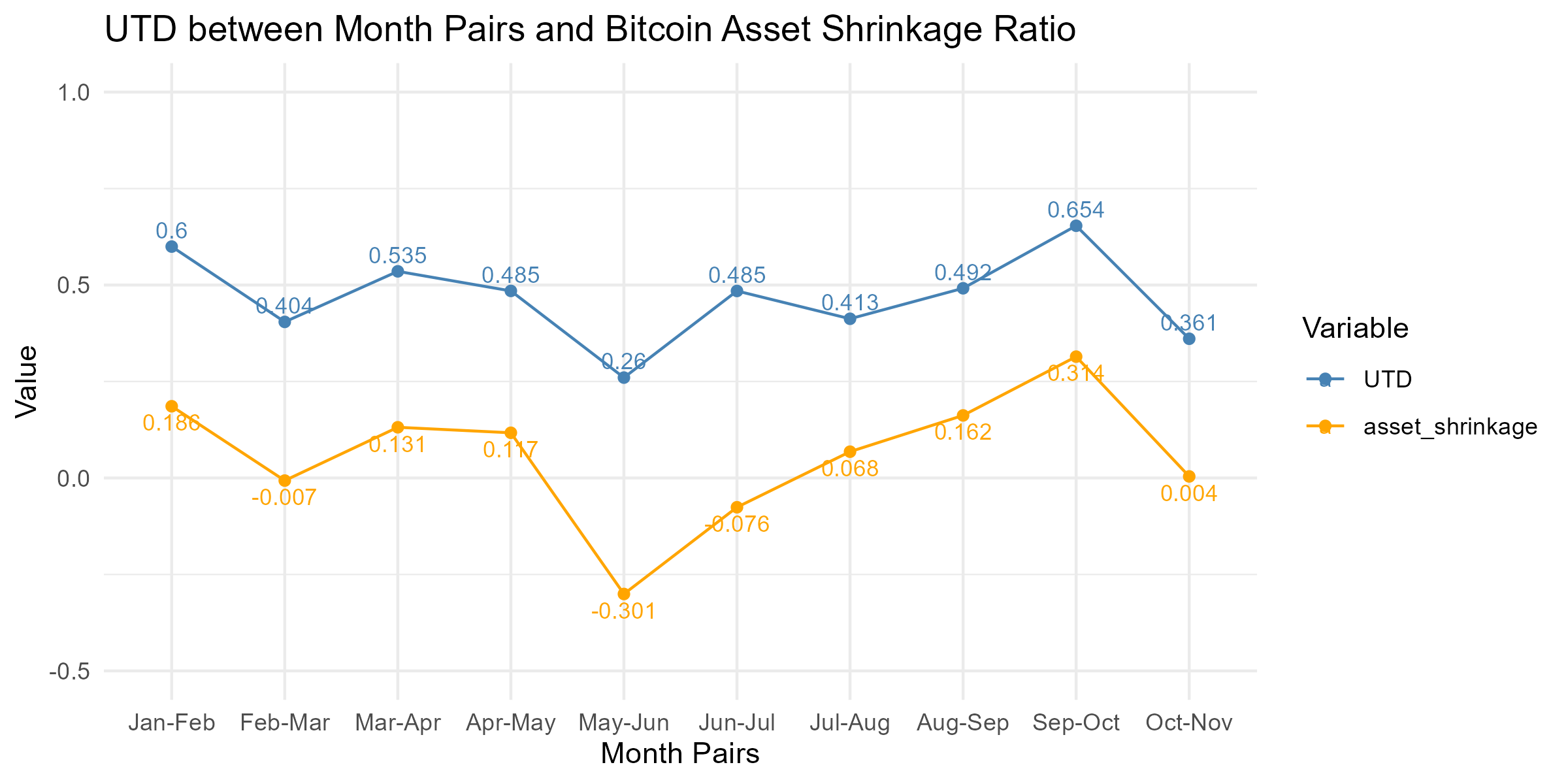}
	\caption{Time series of UTD between month pairs and Bitcoin's asset shrinkage ratio. The fluctuations in UTD and asset shrinkage are highly consistent, suggesting a strong link between user interaction patterns in \textit{BitcoinMarkets} and Bitcoin's market performance. }
	\label{fig: bitcoin}
\end{figure}

Bitcoin price data is sourced from \url{https://cn.investing.com/crypto/bitcoin/btc-usd-historical-data}. Figure~\ref{fig: bitcoin} shows the UTD and the asset shrinkage ratio of Bitcoins over time. We find a strong correlation (0.8849) between UTD in the \textit{BitcoinMarkets} subreddit and Bitcoin's shrinkage ratio, suggesting a strong association between market performance and user engagement.

For example, considering the period of January-February 2014, when Bitcoin's shrinkage ratio was 18\%, UTD was as high as 0.6, meaning that the most active users in January remained active in February. In contrast, May 2014 saw a negative shrinkage ratio (-30\%), and UTD dropped to 0.26, indicating that top contributors in one month were less likely to remain highly active the next. This suggests that when Bitcoin's price falls, engagement among the most active users remains more stable, possibly as they discuss risks and market downturns. Conversely, when prices rise, the set of highly active users changes more, likely due to an influx of new participants. These findings highlight how Bitcoin’s market trends influence user behavior in online financial communities.

\subsection{UTD and seasonal shifts in sports subreddit engagement}
In this section, we analyze the UTD of subreddits \textit{nba} and \textit{CFB} to study the impact of seasonal changes on user interaction patterns. 
%In the online world of sports enthusiasts, the NBA and CFB-related subreddits on Reddit serve as vital platforms for sports fans to communicate and interact. 
%The NBA subreddit mainly centers on discussions about the National Basketball Association. It encompasses a wide range of topics, such as live-game analysis, player trade updates, team strategy discussions, and predictions regarding the championship of each season. Fans on this subreddit share their perspectives on games, evaluations of player performances, and expectations for the future development of teams. The CFB subreddit, on the other hand, focuses on College Football in the United States. The discussion content includes the seasonal performances of various college teams, player recruitment situations, analysis of important game situations, and the competitive landscape within each conference.
\subsubsection{NBA subreddit}

\begin{figure}[h]
	\centering
	\includegraphics[width=0.8\textwidth]{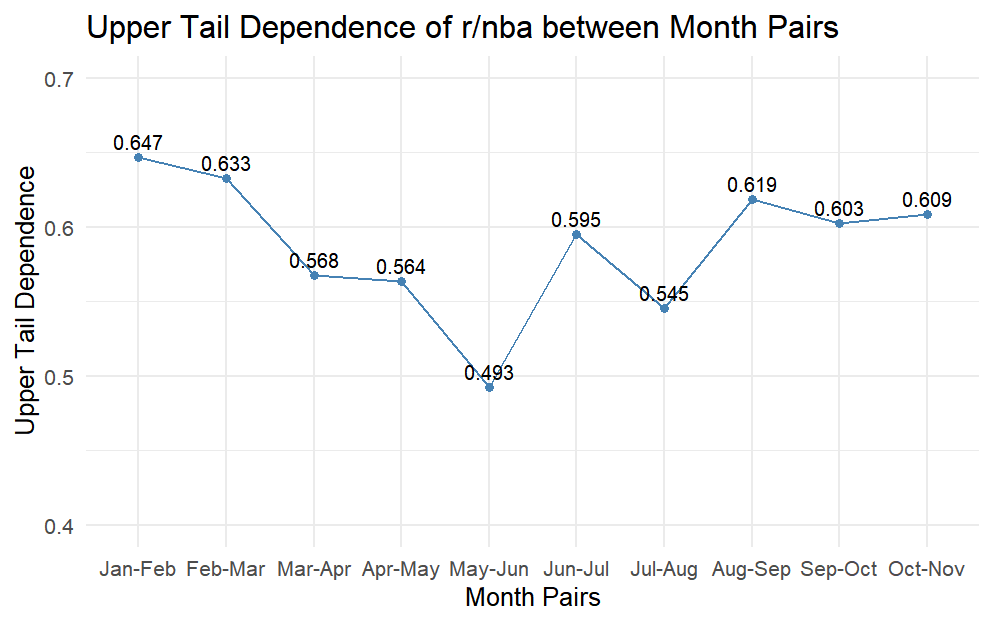}
	\caption{This plot shows the upper tail dependence of the r/nba subreddit's interaction network across consecutive month pairs from January to November. The x-axis indicates month pairs (e.g., "Jan-Feb", "Feb-Mar"), while the y-axis shows the upper tail dependence values, reflecting the degree distribution's dependence in the network.}
	\label{fig:nba-utd}
\end{figure}

The upper tail dependence estimates for the NBA subreddit interaction network, calculated for consecutive months, are plotted in Figure~\ref{fig:nba-utd}. Covering the period from January to November 2014, the analysis yields 10 UTD values. Notably, the May-June UTD value shows a sharp decline, corresponding to the conclusion of the 2014 NBA season on June 16, which aligns with the end of May in our data segmentation. This drop in UTD reflects a shift in user behavior, as engagement among high-degree users decreased after the season ended.

Several factors may explain this decline. During the season, highly engaged users, especially those passionate about their teams, actively participated in online discussions, sharing analyses and predictions. However, after the season’s conclusion, some fans, particularly those whose teams lost, may have disengaged due to disappointment. Additionally, with no ongoing games, there were fewer new topics to sustain high interaction levels. The decline in UTD suggests that a subset of previously active users either reduced their participation or left the subreddit entirely.

Despite this decline, the UTD remained around 0.6, indicating a strong core of loyal users. While interaction frequency decreased, many high-degree users stayed connected to the community, suggesting that engagement would likely rise again with the start of the new season. This pattern highlights the seasonal nature of engagement in sports communities, where interest fluctuates in response to real-world events but remains anchored by a dedicated user base.

\subsubsection{CFB subreddit}

\begin{figure}[h]
	\centering
	\includegraphics[width=0.8\textwidth]{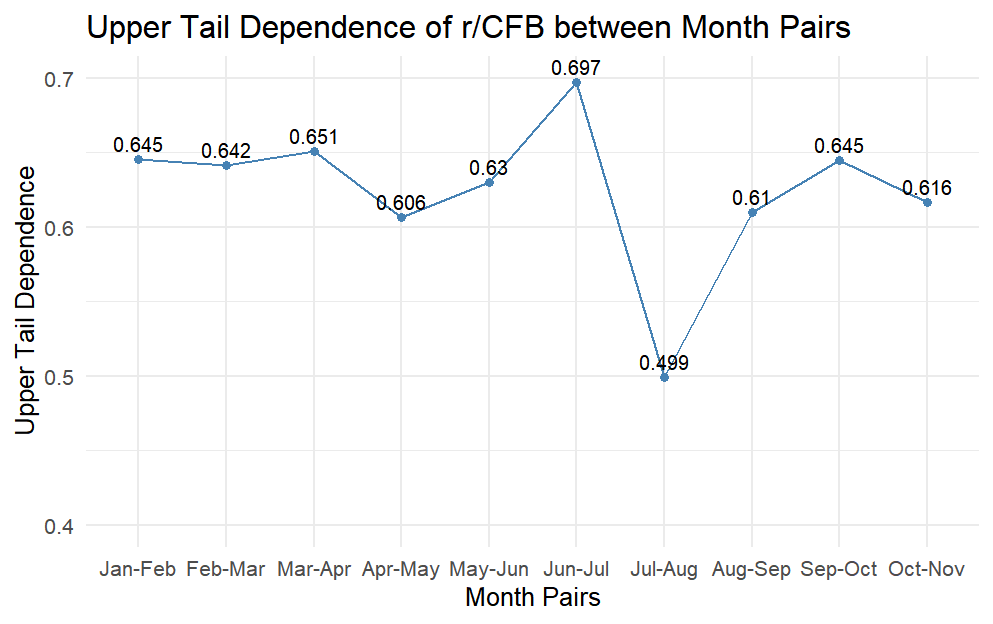}
	\caption{This plot shows the upper tail dependence of the r/CBF subreddit's interaction network across consecutive month pairs from January to November. The x-axis indicates month pairs (e.g., ``Jan-Feb", ``Feb-Mar"), and the y-axis shows the upper tail dependence values, reflecting the dependence in the upper tail of the degree distribution.}
	\label{fig:CBF-utd}
\end{figure}

A similar pattern emerges in the CFB subreddit’s UTD estimates for 2014, as shown in Figure~\ref{fig:CBF-utd}. Among the 10 UTD values, the highest occurs in June-July, followed by a sharp drop in July-August. We note that the 2014 CFB season officially began on August 27, within the eighth month of our data collection.

The June-July peak in UTD suggests that during this part of the off-season, activity was concentrated among a smaller group of highly engaged users. Since discussions at this time may typically focus on preseason rankings, roster changes, and media events, a dedicated group likely contributed to most interactions. However, in July-August, UTD dropped sharply, possibly due to a group of new or returning users as excitement for the season grew, changing the way users interacted.

Once the season started, UTD returned to a more typical level, showing that user engagement had stabilized as discussions shifted to live games. Despite these changes, UTD remained around 0.6 at its lowest point, suggesting a core group of users stayed active over the entire year. This highlights the seasonal nature of sports communities, where engagement rises and falls with the season, but a loyal group of users keeps the discussion going.

\section{Concluding remarks}
\label{sec:conc}
In this paper, we study multilayer inhomogeneous random graphs, focusing on the interlayer dependence of degree distributions in multilayer networks. We establish a connection between the dependence structure of the underlying weight vector $\mathbf{W}$ and the degree distribution, and introduce upper tail dependence as a practical measure to quantify extremal dependence across layers.

Simulation results show that the estimators $\hat{\lambda}(q)$ and $\hat{\lambda}_{D(N)}$ effectively distinguish different asymptotic dependence structures, with accuracy improving as sample size increases. In our real-data analysis of Reddit, we find that seasonal factors strongly influence the upper tail dependence in sports-related subreddits (e.g., ``nba" and ``CFB"), with UTD decreasing at season transitions, reflecting shifts in user engagement. In the BitcoinMarkets subreddit, UTD exhibits a strong correlation with Bitcoin’s monthly asset shrinkage ratio, i.e. higher shrinkage corresponds to greater UTD, indicating more stable engagement among highly active users. This suggests a deep connection between Bitcoin’s market performance and user behavior in financial discussion communities.

Future work may include examining whether similar patterns emerge in other types of online communities or across different social networks. Furthermore, exploring causal relationships between real-world events and network dynamics could provide deeper insights into how external factors shape online interactions.

\section{Acknowledgments}
T. Wang gratefully acknowledges 
Science and Technology Commission of Shanghai Municipality Grant 23JC1400700 and National Natural Science Foundation of China Grant 12301660.
	
	Both authors also thank Shanghai Institute for Mathematics and Interdisciplinary Sciences (SIMIS) for their financial support. This research was partly funded by SIMIS under grant number SIMIS-ID-2024-WE. T. Wang is grateful for the resources and facilities provided by SIMIS, which were essential for the completion of this work.

\appendix
\renewcommand\thetheorem{\Alph{section}.\arabic{theorem}}
\section{Proof of Theorems in Section 3}
\label{app1}
Before proceeding with the proof of Theorem 3.1, we first introduce some existing conclusions that serve as the foundation for our subsequent proof.
\subsection{Fundamental Results of Degree Distribution in MIRG}
From \eqref{def: degree_L1}, we know degrees across nodes are identically distributed. Specifically, Theorem 3.2 in \cite{bhattacharjee2022large} shows that in single-layer networks, large degree behave asymptotically like the weights of associated vertices as follows:
\begin{theorem}[Theorem 3.2 in \cite{bhattacharjee2022large}]
    \label{thm: 3}

    Let \((D_i)_{i = 1}^N\) be the degree sequences of the single-layer graph \(\mathcal{G}_N\) with associated i.i.d. weights \((W_i)_{i = 1}^N\). Define the quantile function \(q(\cdot)\) for weight random variable \(W\) as 
    \[
	q(t) = \inf\{x \ge 0: \mathbb{P}\{W \le x\} \ge 1- 1/t \}, \quad t \ge 1.
    \]
    Denote $\left(t_N\right)_{N \in \mathbb{N}}$ an intermediate sequence that is a positive integer sequence with $t_N<N$ for all $N \in \mathbb{N}, t_N \rightarrow \infty$ and $t_N / N \rightarrow 0$ as $N \rightarrow \infty$, while $\mathbf{1}$ stands for the sequence $(1,1,1, \cdots)$.
    Assume that (A1)-(A4) in \cite{bhattacharjee2022large} are satisfied, then for any \(a > 0\),
    \[
	\lim_{N \to \infty}\frac{1}{t_N}\mathbb{E}\sum_{i = 1}^{N} \mathbf{1}\{D_{i} \ge q(N/t_N) a\ge W_i\} = 0
    \]
    and
    \[
	\lim_{N \to \infty}\frac{1}{t_N}\mathbb{E}\sum_{i = 1}^{N} \mathbf{1}\{W_i \ge q(N/t_N) a\ge D_{i}\} = 0
    \]
\end{theorem}
Theorem~\ref{thm: 3} is a special case obtained from Theorem 3.2 in \cite{bhattacharjee2022large} where the constants all equal to 1, are exactly the values of these constants under MIRG. Also MIRG satisfies assumptions (A1)-(A4) in \cite{bhattacharjee2022large}, which follows from Section 4 in \cite{bhattacharjee2022large} and Lemma 1.1, Lemma 6.5, Lemma 6.6 in \cite{cirkovic2024emergence}.
\subsection{Proof of Theorem 3.1}
    Without loss of generality, we focus on the case of two layers (e.g., layers \(s\) and \(m\)) in the proof, as the results can be naturally extended to multilayer networks by considering pairwise dependencies between any two layers. To simplify notation, we denote \(\hat{\lambda}_{D(N)}^{s,m}\) as \(\hat{\lambda}_{D(N)}\) and \(\lambda_U^{s,m}\) as \(\lambda_U\) throughout the proof. We want to prove that 
    \[
	\hat{\lambda}_{D(N)} \stackrel{p}{\longrightarrow} \lambda_U
    \]
    i.e. 
		\[
		\hat{\lambda}_{D(N)} - \lambda_U \stackrel{p}{\longrightarrow} 0.
		\]
		Define \( w_{l,N} = F_l^{-1}(1-t_N/N), l = 1, 2 \) as the quantile of the weight distribution for layer $l$ where $F_l$ is the cumulative distribution function of $W_l$. The definition of \(\lambda_U\) yields that
		\begin{align*}
			\lambda_U &= \lim_{q \to 1-}\mathbb{P}(W_2>u_2|W_1>u_1)\\
			&= \lim_{N \to \infty}\mathbb{P}(W_2>w_{2,N}|W_1>w_{1,N})\\
			&= \lim_{N \to \infty}\frac{\mathbb{P}(W_1>w_{1,N}, W_2>w_{2,N})}{\mathbb{P}(W_1>w_{1,N})}\\
			&= \lim_{N \to \infty}\frac{N}{t_N} \mathbb{P}(W_1>w_{1,N}, W_2>w_{2,N}).
		\end{align*}
	For the denominator of \(\hat{\lambda}_{D(N)}\), we have 
	\[
	\sum_{i=1}^N \mathbf{1}_{\{D_{i1}(N) > \hat{u}_{1,N}\}} = t_N.
	\]
	Thus we only need to prove that  
		\begin{equation}
				\frac{1}{t_N}\sum_{i=1}^N \mathbf{1}_{\{D_{i1}(N) > \hat{u}_{1,N}\} \cap \{D_{i2}(N) > \hat{u}_{2,N}\}} - \frac{N}{t_N} \mathbb{P}(W_1>w_{1,N}, W_2>w_{2,N})  \stackrel{p}{\longrightarrow} 0.
			\label{eq: proof_main}
		\end{equation}
	The proof of \eqref{eq: proof_main} consists of 3 key steps:\\
		\textbf{Step 1:} Prove that 
		\begin{equation}
			\frac{1}{t_N}\sum_{i=1}^N \mathbf{1}_{\{W_{i1} > w_{1,N}, W_{i2} > w_{2,N}\}} - \frac{N}{t_N}\mathbb{P}(W_{1} > w_{1,N}, W_{2} > w_{2,N}) \stackrel{p}{\longrightarrow} 0 .
			\label{eq: step1}
		\end{equation}
		Define \(A_N = \frac{1}{t_N}\sum_{i=1}^N \mathbf{1}_{\{W_{i1} > w_{1,N}, W_{i2} > w_{2,N}\}}\), the moments satisfying
		\[
		\mathbb{E}[A_N] = \frac{N}{t_N}\mathbb{P}(W_{1} > w_{1,N}, W_{2} > w_{2,N}) < \infty
		\]
		due to the existence of upper tail dependence coefficient \(\lambda_U\) and 
		\begin{align*}
		\text{Var}(A_N)  &= \frac{1}{t_N^2} \sum_{i=1}^N\text{Var}(\mathbf{1}_{\{W_{i1} > w_{1,N}, W_{i2} > w_{2,N}\}})\\
		&= \frac{1}{t_N^2} \sum_{i=1}^N \left(\mathbb{P}(W_{1} > w_{1,N}, W_{2} > w_{2,N}) - [\mathbb{P}(W_{1} > w_{1,N}, W_{2} > w_{2,N})]^2\right)\\
		& \le \frac{1}{t_N^2} \sum_{i=1}^N \left(\mathbb{P}(W_{1} > w_{1,N}, W_{2} > w_{2,N})\right)\\
		& \le \frac{1}{t_N^2} \sum_{i=1}^N \mathbb{P}(W_{1} > w_{1,N})\\
		&= \frac{1}{t_N^2} \sum_{i=1}^N \frac{t_N}{N} =  \frac{1}{t_N} \to 0, \quad \text{ as } N \to \infty.
		\end{align*}
		Then Chebyshev's inequality yields that \(\mathbb{P}(\left|A_N - \mathbb{E}[A_N]\right| > \epsilon) \le \frac{\text{Var}(A_N)}{ \epsilon^2} \le \frac{1}{t_N \epsilon^2} \to 0\), which means that \(A_N - \mathbb{E}[A_N] \stackrel{p}{\longrightarrow} 0\).\\				
	\textbf{Step 2:} Prove that
	\begin{equation}
		\frac{1}{t_N}\sum_{i=1}^N \left(\mathbf{1}_{\{D_{i1}(N) > w_{1,N}\} \cap \{D_{i2}(N) > w_{2,N}\}} - \mathbf{1}_{\{W_{i1} > w_{1,N}, W_{i2} > w_{2,N}\}}\right) \stackrel{p}{\longrightarrow} 0.
		\label{eq: step2}
	\end{equation}
		Note that 
		\begin{align*}
			&\frac{1}{t_N}\left|\sum_{i=1}^N \left(\mathbf{1}_{\{D_{i1}(N) > w_{1,N}, D_{i2}(N) > w_{2,N}\}} - \mathbf{1}_{\{W_{i1} > w_{1,N}, W_{i2} > w_{2,N}\}}\right)\right|\\
			\le&\frac{1}{t_N}\left|\sum_{i=1}^N \left(\mathbf{1}_{\{D_{i1}(N) > w_{1,N}, D_{i2}(N) > w_{2,N}\}} - \mathbf{1}_{\{W_{i1} > w_{1,N}, D_{i2}(N) > w_{2,N}\}}\right)\right|\\
			&+\frac{1}{t_N}\left|\sum_{i=1}^N \left(\mathbf{1}_{\{W_{i1} > w_{1,N}, D_{i2}(N) > w_{2,N}\}} - \mathbf{1}_{\{W_{i1} > w_{1,N}, W_{i2} > w_{2,N}\}}\right)\right|\\
			\le& \frac{1}{t_N}\left|\sum_{i=1}^N \left(\mathbf{1}_{\{D_{i1}(N) > w_{1,N} >W_{i1}, D_{i2}(N) > w_{2,N}\}} + \mathbf{1}_{\{W_{i1} > w_{1,N} > D_{i1}(N), D_{i2}(N) > w_{2,N}\}}\right)\right|\\
			&+\frac{1}{t_N}\left|\sum_{i=1}^N \left(\mathbf{1}_{\{W_{i1} > w_{1,N}, D_{i2}(N) > w_{2,N} > W_{i2}\}} + \mathbf{1}_{\{W_{i1} > w_{1,N}, W_{i2} > w_{2,N} > D_{i2}(N)\}}\right)\right|\\
			\le& \frac{1}{t_N}\left|\sum_{i=1}^N \left(\mathbf{1}_{\{D_{i1}(N) > w_{1,N} >W_{i1}\}} + \mathbf{1}_{\{W_{i1} > w_{1,N} > D_{i1}(N)\}}\right)\right|\\
			&+\frac{1}{t_N}\left|\sum_{i=1}^N \left(\mathbf{1}_{\{D_{i2}(N) > w_{2,N} > W_{i2}\}} + \mathbf{1}_{\{W_{i2} > w_{2,N} > D_{i2}(N)\}}\right)\right|			
		\end{align*}
		From Theorem~\ref{thm: 3}, 
		\[
		\frac{1}{t_N}\sum_{i=1}^N \left(\mathbf{1}_{\{D_{i1}(N) > w_{1,N} >W_{i1}\}} + \mathbf{1}_{\{W_{i1} > w_{1,N} > D_{i1}(N)\}}\right) \stackrel{L_1}{\longrightarrow} 0
		\]
		and
		\[
		\frac{1}{t_N}\sum_{i=1}^N \left(\mathbf{1}_{\{D_{i2}(N) > w_{2,N} > W_{i2}\}} + \mathbf{1}_{\{W_{i2} > w_{2,N} > D_{i2}(N)\}}\right) \stackrel{L_1}{\longrightarrow} 0.
		\]
		Consequently, we have
	\begin{align*}
		&\frac{1}{t_N}\left|\sum_{i=1}^N \left(\mathbf{1}_{\{D_{i1}(N) > w_{1,N} >W_{i1}\}} + \mathbf{1}_{\{W_{i1} > w_{1,N} > D_{i1}(N)\}}\right)\right| \\ 
		& \quad +\frac{1}{t_N}\left|\sum_{i=1}^N \left(\mathbf{1}_{\{D_{i2}(N) > w_{2,N} > W_{i2}\}} + \mathbf{1}_{\{W_{i2} > w_{2,N} > D_{i2}(N)\}}\right)\right| \stackrel{p}{\longrightarrow} 0
	\end{align*}
		and \eqref{eq: step2} is proved.\\
	\textbf{Step 3:} Prove that
	\begin{align}
		\frac{1}{t_N}\sum_{i=1}^N \mathbf{1}_{\{D_{i1}(N) > \hat{u}_{1,N}\} \cap \{D_{i2}(N) > \hat{u}_{2,N}\}} - \frac{1}{t_N}\sum_{i=1}^N \mathbf{1}_{\{D_{i1}(N) > w_{1,N}\} \cap \{D_{i2}(N) > w_{2,N}\}} \stackrel{p}{\longrightarrow} 0.
		\label{eq: step3}
	\end{align}
       Similarly, we have
\begin{align}
	&\left|\frac{1}{t_N}\sum_{i=1}^N \mathbf{1}_{\{D_{i1}(N) > \hat{u}_{1,N}\} \cap \{D_{i2}(N) > \hat{u}_{2,N}\}} - \frac{1}{t_N}\sum_{i=1}^N \mathbf{1}_{\{D_{i1}(N) > w_{1,N}\} \cap \{D_{i2}(N) > w_{2,N}\}}\right|\nonumber\\
	\le & \left|\frac{1}{t_N}\sum_{i=1}^N \mathbf{1}_{\{D_{i1}(N) > \hat{u}_{1,N}\} \cap \{D_{i2}(N) > \hat{u}_{2,N}\}} - \frac{1}{t_N}\sum_{i=1}^N \mathbf{1}_{\{D_{i1}(N) > w_{1,N}\} \cap \{D_{i2}(N) > \hat{u}_{2,N}\}}\right|\nonumber\\
	& + \left|\frac{1}{t_N}\sum_{i=1}^N \mathbf{1}_{\{D_{i1}(N) > w_{1,N}\} \cap \{D_{i2}(N) > \hat{u}_{2,N}\}} - \frac{1}{t_N}\sum_{i=1}^N \mathbf{1}_{\{D_{i1}(N) > w_{1,N}\} \cap \{D_{i2}(N) > w_{2,N}\}}\right|\nonumber\\
	\le & \left|\frac{1}{t_N}\sum_{i=1}^N \mathbf{1}_{\{ \hat{u}_{1,N} < D_{i1}(N) < w_{1,N}, D_{i2}(N) > \hat{u}_{2,N}\}} + \frac{1}{t_N}\sum_{i=1}^N \mathbf{1}_{\{w_{1,N} < D_{i1}(N) < \hat{u}_{1,N}, D_{i2}(N) > \hat{u}_{2,N}\}}\right| \nonumber\\
	& + \left|\frac{1}{t_N}\sum_{i=1}^N \mathbf{1}_{\{D_{i1}(N) > w_{1,N}, \hat{u}_{2,N} < D_{i2}(N) < w_{2,N}\}} + \frac{1}{t_N}\sum_{i=1}^N \mathbf{1}_{\{D_{i1}(N) > w_{1,N},  w_{2,N} < D_{i2}(N) <  \hat{u}_{2,N}\}}\right|\nonumber\\
	\le & \left|\frac{1}{t_N}\sum_{i=1}^N \mathbf{1}_{\{ \hat{u}_{1,N} < D_{i1}(N) < w_{1,N}\}} + \frac{1}{t_N}\sum_{i=1}^N \mathbf{1}_{\{w_{1,N} < D_{i1}(N) < \hat{u}_{1,N}\}}\right| \nonumber\\
	& + \left|\frac{1}{t_N}\sum_{i=1}^N \mathbf{1}_{\{ \hat{u}_{2,N} < D_{i2}(N) < w_{2,N}\}} + \frac{1}{t_N}\sum_{i=1}^N \mathbf{1}_{\{w_{2,N} < D_{i2}(N) <  \hat{u}_{2,N}\}}\right|.\label{eq: step3_1}
\end{align}
To prove \eqref{eq: step3_1}, we first aim to show that \(\frac{1}{t_N}\sum_{i=1}^N \mathbf{1}_{\{ \hat{u}_{1,N} < D_{i1}(N) < w_{1,N}\}} \to 0\) in probability. Let \(Y_N=\frac{1}{t_N}\sum_{i=1}^N \mathbf{1}_{\{ \hat{u}_{1,N} < D_{i1}(N) < w_{1,N}\}}  = \frac{1}{t_N}\sum_{i = 1}^N \mathbf{1}_{\{\frac{D_{i1}(N)}{w_{1,N}} \in (\frac{\hat{u}_{1,N}}{w_{1,N}},1)\}}\). Since
\[\vert Y_N\vert\leq\left\vert Y_N - \nu_{\alpha}\left(\left(\frac{\hat{u}_{1,N}}{w_{1,N}},1\right)\right)\right\vert+\left\vert\nu_{\alpha}\left(\left(\frac{\hat{u}_{1,N}}{w_{1,N}},1\right)\right)\right\vert\]
and 
\[\left\{\vert Y_N\vert>\epsilon\right\}\subseteq\left\{\left\vert Y_N - \nu_{\alpha}\left(\left(\frac{\hat{u}_{1,N}}{w_{1,N}},1\right)\right)\right\vert>\frac{\epsilon}{2}\right\}\cup\left\{\left\vert\nu_{\alpha}\left(\left(\frac{\hat{u}_{1,N}}{w_{1,N}},1\right)\right)\right\vert>\frac{\epsilon}{2}\right\},\] 
by the sub-additivity of probability, we get
\[\mathbb{P}(\vert Y_N\vert>\epsilon)\leq \mathbb{P}\left(\left\vert Y_N - \nu_{\alpha}\left(\left(\frac{\hat{u}_{1,N}}{w_{1,N}},1\right)\right)\right\vert>\frac{\epsilon}{2}\right)+\mathbb{P}\left(\left\vert\nu_{\alpha}\left(\left(\frac{\hat{u}_{1,N}}{w_{1,N}},1\right)\right)\right\vert>\frac{\epsilon}{2}\right).\]
From Proposition 3.5 in \cite{bhattacharjee2022large}, we have
\[
\frac{1}{t_N}\sum_{i = 1}^N \mathbf{1}_{\{\frac{D_{i1}(N)}{w_{1,N}} \in \cdot\}} \stackrel{}{\longrightarrow} \nu_{\alpha}(\cdot) \quad \text { in } \mathbb{M}((0, \infty] )
\]	
where \(\nu_{\alpha}((c, \infty)) = c^{-\alpha}.\) This convergence and inversion [see Proposition 3.2 of \cite{resnick2007heavy}] gives that as $N \rightarrow \infty$, 
\[
\frac{D_{(\lceil t_N y\rceil)1}(N)}{w_{1,N}} \stackrel{p}{\rightarrow} y^{-1 / \alpha}. 
%\quad \text { in } D(0, \infty] .
\]
In particular, as $N \rightarrow \infty$,
$$
\frac{D_{(\lceil t_N \rceil)1}(N)}{w_{1,N}} \stackrel{p}{\rightarrow} 1.
$$
Here, \(D_{(\lceil k\rceil)1}(N)\) is the \(\lceil k\rceil\)-th order statistic of the degree sequences \(\{D_{i1}(N)\}_{i = 1}^N\) of layer 1, arranged in descending  order. Specifically,
\[
\hat{u}_{1,N} = D_{(\lceil t_N \rceil)1}(N),
\]
which implies \( \frac{\hat{u}_{1,N}}{w_{1,N}}\stackrel{p}{\rightarrow} 1.\) For an arbitrary \(\epsilon > 0\), the definition of convergence in probability implies that there exists \(N_1\) such that for all \(N > N_1\),
\[
\mathbb{P}\left(\left|\frac{\hat{u}_{1,N}}{w_{1,N}} - 1\right| < \epsilon\right) > 1 - \epsilon.
\]
Furthermore, by the continuity of the measure \(\nu_\alpha\) at \(1\), there exists \(\delta > 0\) such that whenever \(\left|\frac{\hat{u}_{1,N}}{w_{1,N}} - 1\right| < \delta\), the measure of the interval satisfies,
\[
\nu_\alpha\left(\left(\frac{\hat{u}_{1,N}}{w_{1,N}}, 1\right)\right) < \frac{\epsilon}{2}.
\]
Applying the convergence in probability again to this \(\delta\), there exists \(N_2\) such that for all \(N > N_2\),
\[
\mathbb{P}\left(\left|\frac{\hat{u}_{1,N}}{w_{1,N}} - 1\right| < \delta\right) > 1 - \frac{\epsilon}{2}.
\]
Combining these results, for \(N > \max\{N_1, N_2\}\), we have
\[
\mathbb{P}\left(\left|\nu_\alpha\left(\left(\frac{\hat{u}_{1,N}}{w_{1,N}}, 1\right)\right)\right| > \frac{\epsilon}{2}\right) \leq \mathbb{P}\left(\left|\frac{\hat{u}_{1,N}}{w_{1,N}} - 1\right| \geq \delta\right) < \frac{\epsilon}{2}.
\]
Meanwhile, since \(\frac{1}{t_N}\sum_{i = 1}^N \mathbf{1}_{\{\frac{D_{i1}(N)}{w_{1,N}} \in \cdot\}} \stackrel{}{\longrightarrow} \nu_{\alpha}(\cdot)\), for the given \(\frac{\epsilon}{2}>0\), there exists \(N_3\) such that for \(N > N_3\),
\[P\left(\left\vert Y_N - \nu_{\alpha}\left(\left(\frac{\hat{u}_{1,N}}{w_{1,N}},1\right)\right)\right\vert>\frac{\epsilon}{2}\right)<\frac{\epsilon}{2}.\]
Let \(N_0=\max\{N_1,N_2,N_3\}\). For \(N > N_0\), we have
\[P(\vert Y_N\vert>\epsilon)<\epsilon,\]
i.e. \(\frac{1}{t_N}\sum_{i = 1}^N \mathbf{1}_{\{\hat{u}_{1,N} < D_{i1}(N) < w_{1,N}\}} \stackrel{p}{\to} 0\). \\
Similarly, we can prove that 
%\(\frac{1}{t_N}\sum_{i = 1}^N \mathbf{1}_{\{w_{1,N} < D_{i1}(N) < \hat{u}_{1,N}\}}\), \(\frac{1}{t_N}\sum_{i = 1}^N \mathbf{1}_{\{\hat{u}_{2,N} < D_{i2}(N) < w_{2,N}\}}\) and \(\frac{1}{t_N}\sum_{i = 1}^N \mathbf{1}_{\{w_{2,N} < D_{i2}(N) < \hat{u}_{2,N}\}}\) 
the other 3 terms of the righthand side of \eqref{eq: step3_1} all converge to \(0\) in probability. This completes the proof of \textbf{Step 3}, and thus the overall proof of the theorem is established. \(\hfill \square\)

\bibliographystyle{elsarticle-num} 
\bibliography{Bibliography-MM-MC}



\end{document}